\documentclass[%
 reprint,
 amsmath,amssymb,
 aps,
pra,
]{revtex4-2}

\usepackage{graphicx}% Include figure files
\usepackage{dcolumn}% Align table columns on decimal point
\usepackage{bm}% bold math
\usepackage{algorithmic}
\usepackage{graphicx}
\usepackage{textcomp}
\usepackage{varwidth}
\usepackage{xcolor}
\usepackage{quantikz}
\usepackage[colorlinks=true,bookmarks=false,citecolor=red,urlcolor=red]{hyperref} 

\usepackage{enumitem}

\usepackage{float}

\begin{document}

\preprint{APS/123-QED}
%An architecture for distributed fault tolerant quantum computing with finite resources on trapped-ion systems with  spatial and temporal mode
%multiplexing for enhanced entanglement generation 

\title{Multiplexed bi-layered realization of fault-tolerant quantum computation \\over optically networked trapped-ion modules}% Force line breaks with \\

\author{Nitish Kumar Chandra$^{1}$}
\email{nkc16@pitt.edu}
\author{Saikat Guha$^{2,3}$}
\email{saikat@umd.edu}
\author{Kaushik P. Seshadreesan$^{1}$}
\email{kausesh@pitt.edu}
\affiliation{$^{1}$Department of Informatics and Networked Systems, School of Computing and Information, University of Pittsburgh, Pittsburgh, PA 15213, USA}
\affiliation{$^{2}$Wyant College of Optical Sciences, University of Arizona, Tucson, AZ 85721, USA}
\affiliation{$^{3}$Department of Electrical and Computer Engineering, A. James Clark School of Engineering, University of Maryland, College Park, MD 20742, USA}

\begin{abstract}
We study an architecture for fault-tolerant measurement-based quantum computation (FT-MBQC) over optically-networked trapped-ion modules. The architecture is implemented with a finite number of modules and ions per module, and leverages photonic interactions for generating remote entanglement between modules and local Coulomb interactions for intra-modular entangling gates. We focus on generating the topologically protected Raussendorf-Harrington-Goyal (RHG) lattice cluster state, which is known to be robust against lattice bond failures and qubit noise, with the modules acting as lattice sites. To ensure that the remote entanglement generation rates surpass the bond-failure tolerance threshold of the RHG lattice, we employ spatial and temporal multiplexing. For realistic system timing parameters, we estimate the \textit{code cycle time} of the RHG lattice and the ion resources required in a bi-layered implementation, where the number of modules matches the number of sites in two lattice layers, and qubits are reinitialized after measurement. For large distances between modules, we incorporate quantum repeaters between sites and analyze the benefits in terms of cumulative resource requirements. Finally, we derive and analyze a qubit noise-tolerance threshold inequality for the RHG lattice generation in the proposed architecture that accounts for noise from various sources. This includes the depolarizing noise arising from the photonically-mediated remote entanglement generation between modules due to finite optical detection efficiency, limited visibility, and presence of dark clicks, in addition to the noise from imperfect gates and measurements, and memory decoherence with time. Our work thus underscores the hardware and channel threshold requirements to realize distributed FT-MBQC in a leading qubit platform today---trapped ions.
\end{abstract}

\maketitle
\section{Introduction}
\label{Intro}

% Over the past decade, notable progress has been made in conceptualizing scalable, fault-tolerant quantum computing architectures~\cite{PhysRevX.4.031022,PhysRevA.89.022317,roh2023generation}. However, realizing them with current technology remains challenging. 
% An effective architecture must not only deploy quantum error-correcting codes capable of sustaining  high noise and loss error thresholds to ensure low error probability per logical gate, but must also specify implementation of a universal set of logical gates, must have low spatial and temporal overheads, and must process error information quickly in tandem with the quantum computation~\cite{gottesman2022opportunities, Campbell2017}.

Over the past decade, significant progress has been made in conceptualizing scalable, fault-tolerant quantum computing architectures~\cite{PhysRevX.4.031022,PhysRevA.89.022317,roh2023generation}. However, realizing these architectures with current technology remains a formidable challenge. An effective architecture must not only employ quantum error-correcting codes that can withstand high loss and noise error thresholds to maintain a low error probability per logical gate, but must also enable the implementation of a universal set of logic gates. Additionally, it should have minimal spatial and temporal overheads, and process error information efficiently in parallel with the quantum computation~\cite{gottesman2022opportunities, Campbell2017}

%To integrate error-corrected qubits and execute complex quantum algorithms, it is crucial to address error minimization and qubit loss. Effective architectures must have a high noise threshold to ensure low error probability per logical gate and support a universal set of logical gates~\cite{Campbell2017}. These challenges are further compounded by experimental limitations and environmental noise, restricting the number of usable physical qubits~\cite{gottesman2022opportunities}. 

A widely pursued approach to quantum computing is measurement-based quantum computation (MBQC)~\cite{PhysRevA.71.062313, RAUSSENDORF20062242,Raussendorf_2007}, where the computation proceeds through single-qubit measurements on a large entangled resource cluster state with feedforward. Fault-tolerant (FT)-MBQC robust to qubit loss and noise errors can be achieved using 3D cluster states, where logical qubits are encoded, e.g., in topologically degenerate states of non-Abelian anyons, and manipulated through braiding~\cite{KITAEV20032}. These 3D cluster states can be realized in bosonic systems such as optical modes~\cite{PhysRevResearch.2.023138,Du:23}, and matter-based systems such as trapped ions~\cite{PhysRevLett.102.170501} and negatively charged nitrogen vacancy (NV$^{-}$) centers in diamond~\cite{PhysRevX.4.031022}. The main challenge lies in realizing large multi-qubit systems with accurate control~\cite{8993497} to host the cluster states. A viable option to tackling this problem is to optically link small few-qubit modules to scale up the system~\cite{10.1145/3311879}. Cluster state generation in such a modular distributed setting where the modules occupy the cluster lattice sites typically involves generating remote entanglement between qubits across modules and connecting them via local entangling gates and measurements~\cite{PhysRevA.76.062323}. 

Along these lines, Monroe et al.~\cite{PhysRevA.89.022317} proposed a modular universal scalable ion trap quantum computer (MUSIQC) architecture, and showed that fault-tolerant quantum computing %over large distances 
can be achieved in small-scale trapped ion modules in the presence of both fast and slow remote entanglement generation between modules. 
The architecture involves creating the 3D cluster state called the Raussendorf-Harrington-Goyal lattice~\cite{RAUSSENDORF20062242} (up to local Hadamard gates on the edge qubits) by preparing Bell states between trapped ion modules (qubits therein) present at the lattice sites along the lattice edges and fusing them at the modules by performing CNOT gates followed by local Pauli measurements. The process of generating a unit cell of a 3D cluster state can be extended along all three spatial axes to construct as large a cluster state as required to perform large-scale quantum computations over multiple logical qubits and of arbitrary circuit depth. Ref.~\cite{PhysRevLett.102.170501} demonstrated that the RHG lattice cluster state can in fact be realized with a 2D hexagonal array of trapped ion modules. This array unfolds into two layers of the RHG lattice, which, when reinitialized and measured repeatedly, form the RHG lattice cluster state over time, removing the need for spatial construction. Quantum computation then proceeds by teleporting information between these two layers. This approach has been analyzed for quantum computing over NV$^{-}$ centers in diamond in Ref.~\cite{PhysRevX.4.031022}.

% Remote entanglement generation is evidently a critical element of the MUSIQC architecture. Being enabled via ion-photon entangled pair generation followed by probabilistic linear optical Bell statement measurement over a pair of such photons for entanglement swapping, which results in an ion-ion Bell pair, the process is inherently stochastic~\cite{10.5555/2011617.2011618}. Failures result in a defective RHG lattice with missing bonds, which in the absence of measurement errors can tolerate upto 6.5 \% heralded bond failure rate~\cite{PhysRevLett.113.140403} (heralded here implies the locations of missing bonds are known). An adaptive scheme, where qubits are measured in an alternative basis~\cite{PhysRevLett.113.140403} improves this threshold to 14.5 \%, alleviating the adverse impact of missing bonds~\cite{PhysRevA.97.030301}. 
% However, remote Bell pair generation via passive linear optics and unentangled ancillae typically incurs upwards of 25 \% bond failure rate~\cite{PhysRevLett.113.140403}. 

Remote entanglement generation is a crucial step in realizing the MUSIQC architecture. It is achieved via ion-photon entangled pair generation, followed by probabilistic linear optical Bell-state measurements over a pair of such photons for entanglement swapping, resulting in an ion-ion Bell pair. This process is inherently stochastic~\cite{10.5555/2011617.2011618}. Failures lead to a defective RHG lattice with missing bonds, which, in the absence of measurement errors, can tolerate up to a $6.5\%$ heralded bond failure rate~\cite{PhysRevLett.113.140403} (where heralded means that the locations of the missing bonds are known) in order for the system to remain fault tolerant. An adaptive scheme, in which qubits are measured in an alternative basis, improves this threshold to $14.5\%$, mitigating the adverse impact of missing bonds~\cite{PhysRevA.97.030301}. However, remote Bell pair generation via passive linear optics and unentangled ancillae typically results in bond failure rates exceeding $25\%$~\cite{PhysRevLett.113.140403}. Practical implementations of photonically-mediated remote entanglement generation are typically also noisy, which add to the noise from other sources such as gate and measurement infidelities and memory decoherence with time, on the RHG lattice qubits. In the absence of bond failure errors, the overall qubit noise tolerance threshold of the RHG lattice was derived in Ref.~\cite{PhysRevA.89.022317}.

In this work, we consider a bi-layered realization of the RHG lattice akin to the architecture of Ref.~\cite{PhysRevLett.102.170501}. We incorporate spatial and time multiplexing to boost the success rate of remote Bell pair generation between modules. We determine the multiplexing and the associated ion resource requirements at the modules to operate within the bond failure tolerance threshold for FT-MBQC, including the case where the channel between modules is interspersed with quantum repeaters. We consider a realistic model for photonic interactions in remote entanglement generation based on single-photon dual-rail qubits~\cite{PhysRevResearch.5.033149}, which produces noisy Bell pairs between ions across modules. We derive a noise-tolerance threshold inequality that factors in noise arising from the optical detection process in remote entanglement generation, along with noise from gate and measurement infidelities, and memory decoherence with time. We analyze the inequality over the space of the various relevant system parameters to identify valid operating regions in the parameter space for FT-MBQC. %in relation to the noise threshold of the RHG lattice thus constructed. %on the noise levels of the generated remote entanglement 
%and incorporating insights from the other aforementioned prior works, here we study remote Bell pair generation with spatial and time mutliplexing. %and therefore more capable of attaining the loss tolerance threshold for the adaptive scheme than the time-only multiplexed scheme of the MUSIQC architecture. 
%We also explore the noise tolerance of the architecture in relation to a well-defined optical scheme for remote entanglement generation based on mediation by dual-rail optical qubits under realistic 

More precisely, the main contributions of our work include the following:
\begin{itemize}
   \item We outline the approach to generate the RHG lattice with just as many trapped ion modules as the number of sites in two layers of the lattice based on re-initializing the qubits therein post measurement in MBQC and space-and-time multiplexed remote entanglement generation between modules.
    \item We derive the time required to complete one \textit{code cycle} in FT-MBQC, accounting for various system timing parameters. Additionally, we calculate the ion resources required at the modules to support the multiplexed remote entanglement generation at a failure rate below the maximum bond failure tolerance rate threshold of the RHG lattice.%, i.e., corresponding to the absence of noise errors.
    \item To address large distances between modules, we incorporate quantum repeaters and evaluate its impact on performance, as well as the associated ion resource requirements.
    
    % In order to address large distances between modules, we consider quantum repeaters and determine the resulting repeater-enhanced performance and corresponding resource requirements. 
    % We repeat our analysis for different numbers of quantum repeaters positioned between the two nodes. 
    \item We consider noisy remote entanglement between modules, represented by two-qubit mixed entangled states, generated by considering realistic and imperfect single-photon dual-rail qubit-mediated interactions. The infidelity arises from factors such as limited channel transmissivity, detector noise, and dark clicks. Combining it with noise from imperfect gates and measurements, and memory decoherence over the course of generation of the RHG lattice, and comparing the overall noise to the maximum noise tolerance threshold of the RHG lattice, we derive a threshold inequality. Using the inequality, we identify valid regions in the parameter space where the threshold is met.
    
    % We consider noisy remote entanglement between modules in the form of two qubit mixed entangled states prepared by realistic and imperfect photonic qubit interactions, where the infidelity arises from finite channel transmissivity, presence of detector noise and detector dark clicks. By weighing the noise in the state with the maximum noise tolerance rate threshold of the RHG lattice, i.e. corresponding to the absence of bond-failure errors, we determine the condition involving various   parameters to remain fault tolerant and find region of various system parameters to operate the scheme within the threshold.
\end{itemize}

In Sec.~\ref{trapped ion}, we briefly review ion trap quantum computing. In Sec.~\ref{tcqc}, we discuss FT-MBQC, also referred to as topological cluster state quantum computation (TCQC) and its two layer realization. In particular, we discuss the RHG lattice cluster state as an illustrative example of a topologically protected cluster state, its fault tolerance, and bi-layered realization. 
In Sec.~\ref{generation}, we discuss the bi-layered realization of the RHG lattice cluster state over optically networked trapped ion modules, and the use of spatial and time multiplexed remote entanglement generation between modules to deal with bond failures. %We show that finite ion resources at the two layers of trapped ion modules are sufficient. 
% In Section~\ref{repeater}
Here, we also discuss the role of quantum repeaters when the lattice sites (modules) are widely separated. In Sec.~\ref{fault}, for the same setting of optically networked trapped ion modules, we consider a realistic model for the photon-mediated remote entanglement generation resulting in imperfect Bell pairs between modules. We discuss the implications therein due to the noise tolerance rate threshold of the RHG lattice. We conclude in Section~\ref{conc} with a summary.

\section{Ion trap Quantum Computing} \label{trapped ion}
Trapped ions have emerged as a promising platform for quantum computing, with well-defined qubits having long coherence times~\cite{doi:10.1126/sciadv.1601540,doi:10.1073/pnas.1618020114}, the ability to perform universal logic gates, and efficient state initialization and measurement~\cite{Amini_2010}. These features make them an excellent candidate for scalable fault-tolerant quantum computation. 
Trapped ion systems consist of atomic ions like $^{171}\textrm{Yb}^{+}$ and $^{40}\textrm{Ca}^{+}$ that are confined between electrodes by AC and DC electric fields. The ions encode qubits in their stable internal spin states. These spin states can be either the hyperfine ground states or a mix of ground and excited optical states. 
Implementing logical gates involves precisely directing tuned laser beams to the ions, inducing coupling between qubit states and resulting in transitions among spin states~\cite{10.1063/1.5088164}. For single-qubit gates enabling arbitrary rotations, a laser beam is accurately targeted at the ion, causing a change in its internal state through Raman Transitions. To apply two-qubit gates, lasers are used to induce coulombic interactions between pairs of ions, facilitating the alteration of one ion's spin state influenced by the state of another. To enable multi-qubit operations, multiple ions can be trapped to form a chain, sharing a common vibrational mode that acts as a platform for executing complex multi-qubit gates  ~\cite{ahsan2015architecture}.

%\subsection{Architectures}
Various architectural designs have been proposed for trapped ion quantum computing, each striving to achieve scalability~\cite{Kielpinski2002,10.1109/MICRO.2005.9,PhysRevA.89.022317,10.1145/3311879}. The main challenge lies in effectively coupling distant qubits by establishing robust communication pathways between remote sites. %, which is pivotal for devising a viable quantum processor architecture. 
Among such designs is the Quantum Charge-Coupled Device (QCCD) architecture, pioneered by Kielpinski et al.~\cite{Kielpinski2002}, which segregates the processor into discrete memory and interaction regions, thereby augmenting scalability and orchestrating a coherent data flow within the system. Another noteworthy approach is the Quantum Logic Array (QLA) architecture introduced by Metodi et al., which harnesses surface traps arranged into short arrays of ion chains on a 2-D planar surface~\cite{10.1109/MICRO.2005.9}. These microarchitectures exhibit variations in their arrangement of computational zones, ancilla generation sections, memory compartments for idle qubits, and teleportation network resources~\cite{10.1145/3311879}. They also facilitate fundamental operations such as ion shuttling that are critical for enabling joint manipulations of distance qubits.

While QCCD and QLA architectures show promise for small to medium-scale systems, scaling them up to millions of qubits remains a major challenge, especially when it comes to connecting distant qubit regions. The exact method of connection depends on the processor design, but the challenges remain the same. For instance, moving ions between trapping zones using electromagnetic fields or lasers introduces difficulties. As the number of qubits increases, so does the complexity of the interconnects, which can lead to problems like more crosstalk, higher error rates, and difficulty maintaining coherence and entanglement over long distances. The diffraction of optical beams used to control ions adds to these challenges, limiting scalability. Current research is working to overcome these obstacles by improving ion trapping and manipulation techniques, developing new materials and fabrication methods for more precise connections, and integrating advanced control and error-correction methods into the system~\cite{10.1063/1.5088164}.

% Though the QCCD and QLA architectures are promising for small to medium-scale systems, scaling up to millions of qubits and more still poses a formidable challenge especially with regard to interconnecting distant regions of qubits. The modes of connections may vary based on the processor's specific design and setup, but the challenge remains all the same. For instance, difficulties arise in moving ions between trapping zones using electromagnetic fields or lasers to enable quantum operations. With an increased interconnect complexity due to growing qubit count, there may also be problems like heightened crosstalk, increased error rates, and difficulties in maintaining coherence and entanglement over longer distances. The diffraction of optical beams used for ion control can further complicates interconnects, constraining scalability. Ongoing research focuses on enhancing interconnectivity through advancing ion trapping and manipulation techniques, pioneering innovative materials and fabrication methods to ensure precise interconnects, and integrating sophisticated control and error-correction mechanisms into the architecture of the system~\cite{10.1063/1.5088164}.
\begin{figure}%[H]
       \centering
\includegraphics[width=.85\columnwidth]{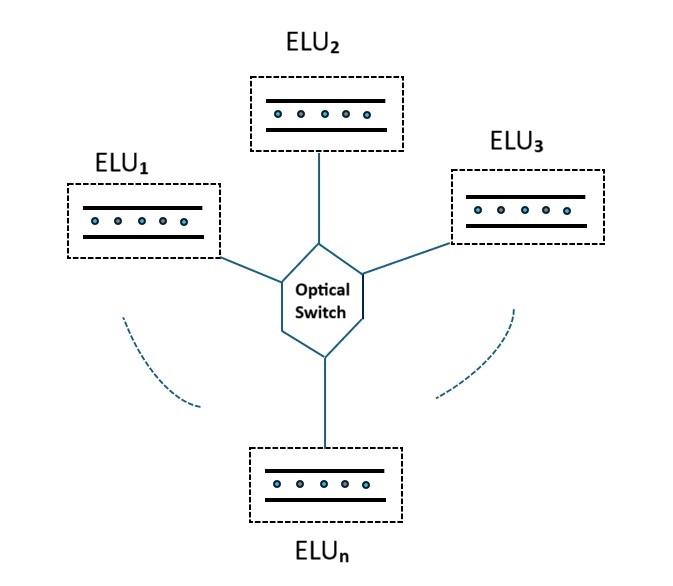}
        \caption{ Illustration of the MUSIQC architecture, consisting of $N$ ELUs that are interconnected via an optical switch.}
         \label{musiqc}
    \end{figure}
\begin{figure}%[H]
       \centering   
\includegraphics[width=.90\columnwidth]{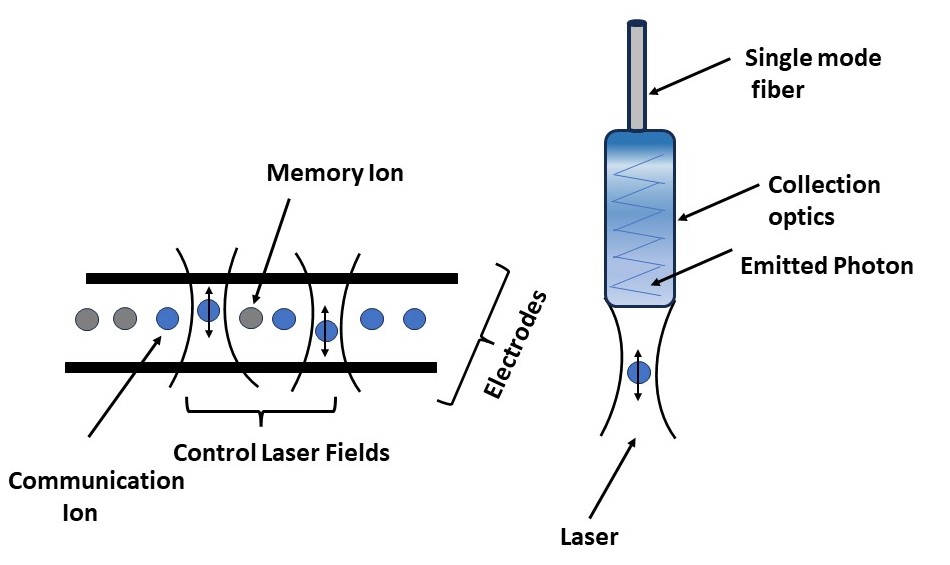}
        \caption{ Essential components of an Elementary Logic
Unit (ELU). It comprises of 50–100 trapped ions.}
         \label{ELU}
    \end{figure}
\begin{figure}%[H]
       \centering   
\includegraphics[width=.99\columnwidth]{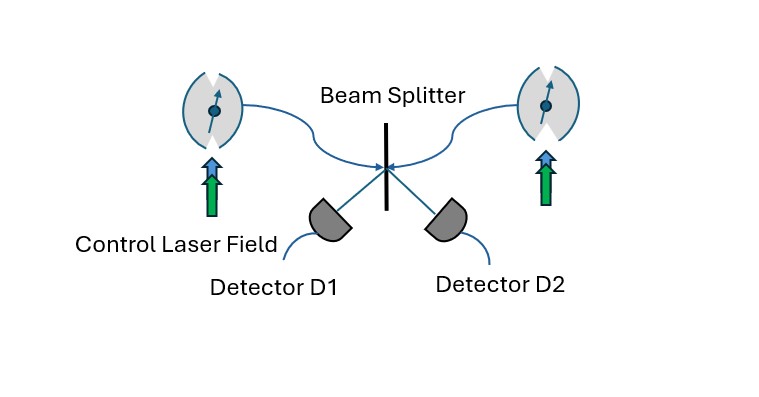}
        \caption{ Schematic illustration of probabilistic entanglement generation between two ions situated in separate Elementary Logical Units (ELUs). The configuration includes detectors D1 and D2, along with a beam splitter. Entanglement generation is heralded by distinct detector click patterns, enabling the identification of successful entangled pairs.}
         \label{Bell}
    \end{figure}
    
To overcome the limitations of the QCCD architecture, Monroe et al. introduced the MUSIQC architecture (see Fig.~\ref{musiqc}), a modular universal scalable ion trap quantum-computer architecture. It involves Elementary Logic Units (ELUs) consisting of 50–100 trapped ions each, a reconfigurable optical switch, and beamsplitters and photodetectors. More precisely, an ELU (see Fig.~\ref{ELU}) consists of \textit{dual-species trapped-ions (DSTI)}, namely, \textit{memory} and \textit{communication} ions. The memory ions are chosen to be a species with long qubit coherence time so as to serve as effective information storage and for local quantum processing, e.g., Yb$^+$. The communication ions are chosen to be with good optical properties so as to enable efficient inter-ELU communications, e.g., Ba$^+$. The latter are linked to a photonic interface and upon excitation and deexcitation using lasers can generate ion-photon entanglement involving the polarization or frequency degree of freedom of the photon. The generated photons when collected and propagated using optical fiber to the re-configurable optical switch, can be used to connect the different ELUs. For example, an ion-ion remote Bell pair can be generated between a pair of ELUs via optical Bell state measurements performed over the photons using beamsplitters and photodetectors (See Fig.~\ref{Bell}). The Bell pairs between ELUs can then be used to execute two-qubit quantum logic operations between any pair of physical qubits present across the ELUs. 

Building upon the MUSIQC architecture, Ahsan et al.~\cite{ahsan2015optimization} adapted it with a modification, namely, the integration of the Quantum Logic Array (QLA) architecture within each ELU instead of the initially envisioned linear ion arrangement. The QLA architecture brings specific benefits, including enhanced gate fidelities and reduced error rates in quantum operations, thereby optimizing the computational efficiency of the system.

\section{Fault-Tolerant Measurement-based Quantum Computation}\label{tcqc}
Fault Tolerant Measurement-Based Quantum Computation (FT-MBQC), also known as Topological Cluster State Quantum Computation (TCQC)~\cite{PhysRevA.71.062313, RAUSSENDORF20062242,Raussendorf_2007}, involves the use of a resource cluster state defined on a three-dimensional lattice and single-qubit measurements and operations. The computation advances by sequentially measuring cross-sectional layers of the lattice (parallel to the $x-y$ plane) that encode error-corrected logical qubits, with the transformed logical information getting teleported between layers via the bonds in the cluster along the third ($z$) dimension.

A cluster state can be conveniently described using the stabilizer formalism as the simultaneous $+1$ eigenstate of the stabilizer generators
\begin{align}
   K_{a}=X_{a} \bigotimes_{b \in n(a)} Z_b, 
\end{align}
where $a$ spans the sites of any lattice configuration that the qubits are arranged in, $n(a)$ is the set of nearest neighbors of site $a$ in the lattice, and $X$ and $Z$ are the $2 \times 2$ Pauli operators. 
One way to generate a cluster state is to arrange qubits prepared in the $|+\rangle$ state at the lattice sites and applying controlled-Z (CPhase) gates along all the bonds in the lattice. %, resulting in a configuration of intertwined primal and dual cubic lattices crucial for error correction \kaushik{is the latter generic or true only for RHG lattice?}. 
When the lattice sites are present over distinct modules, e.g., trapped-ion ELUs, bonds of the lattice can be generated by distributing Bell states between the modules followed by local measurements (MUSIQC architecture). 
%\kaushik{I modified this a bit. I think the physical coordinates don't convey any information; so may also omit it. Do you agree? If not, we can bring this back. I think the RHG lattice structure with stabilizers could be added here.}

\begin{figure}%[H]
       \centering
\includegraphics[width=0.75\columnwidth]{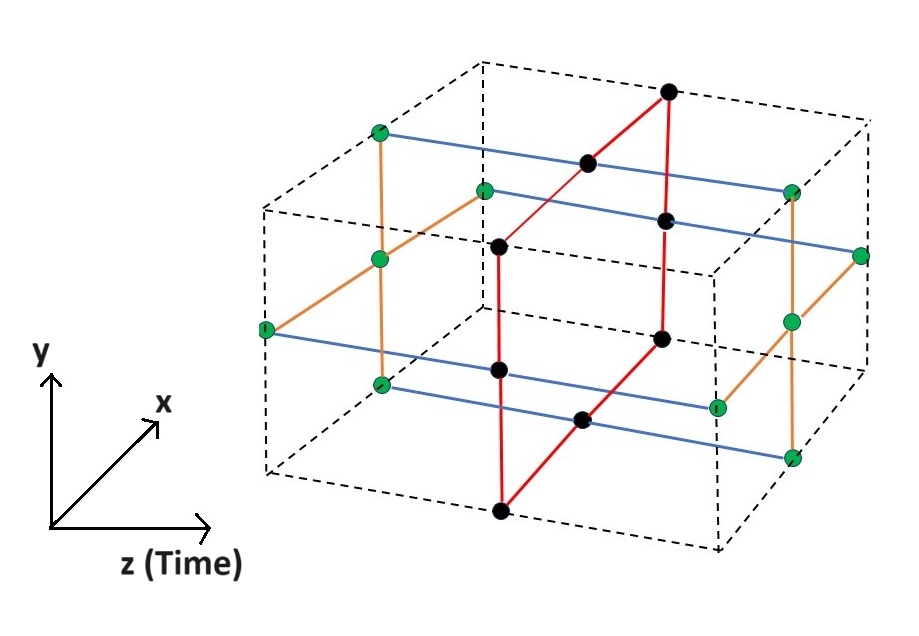}
        \caption{ A ($2 \times 2 \times 2$) unit cell of the Rausendorff-Harrington-Goyal (RHG) lattice. The cell contains sites at each edge and face center of the 3D cubic lattice, represented as green and black colors, respectively. Two different kinds of layers exist within the lattice, marked by green and black sites. Black sites are connected by red bonds, while green sites are linked by orange bonds. Blue bond connects the inter-layer bonds. The faces of the cells are referred to as either primal or dual.}
        \label{fig:RHG_schematic}
    \end{figure}

\subsection{Rausendorff-Harrington-Goyal (RHG) Lattice}
The 3D lattice of interest in this work is the Rausendorff-Harrington-Goyal (RHG) lattice (see Fig.~\ref{fig:RHG_schematic}), which has it sites at all the face and edge centers of the 3D cubic lattice. 
The RHG lattice is  an adaptation of surface-code into the MBQC framework, where alternating primal and dual layers of a surface code are~\textit{foliated} along a third dimension~\cite{bolt2016foliated}. 
It can also be viewed as a pair of interleaved cubic lattices that are dual to one another, namely, primal lattice of all face-centered qubits alone and dual lattice of all edge-centered qubits alone, respectively.
The stabilizer operators of a unit cell volume $\partial q$ of the primal lattice, e.g., are given by %\kaushik{can you please check what about stabilizers with X on edge centers? Do they cancel off?}
\begin{align}
   \{K_f| \ \ K_f=X_{f} \bigotimes_{e \in \partial f} Z_e, \ f\in\partial q\}, 
\end{align}
such that their product simplifies to
\begin{align}
    K_{\partial q}&=\prod_{f\in\partial q} K_f=\prod_{f\in\partial q} X_f.\label{RHG_unit_stabilizer}
\end{align}
The operator $K_{\partial q}$ thus corresponds to measuring each face centered qubit in the $X$ basis. 
It is termed a check operator as it serves as a parity check for the unit cell. 

\subsubsection{Bond Failure Tolerance} 

% The RHG lattice exhibits robustness against bond failure errors, and in the absence of measurement errors, can sustain up to $14.5$ percent bond failure rate when an adaptive measurement scheme is used

% The RHG lattice demonstrates significant robustness against bond failure errors, tolerating up to a $14.5$ percent bond failure rate with an adaptive measurement scheme in the absence of measurement errors. To handle missing bonds, two strategies are employed: a non-adaptive scheme that requires no additional quantum processing and supports a $6.5$ percent bond-loss rate, and an adaptive scheme that measures qubits in an alternative basis to eliminate them, thereby enhancing threshold performance~\cite{PhysRevA.97.030301}. 
% This is based on an adaptive scheme, where bond failures are associated with qubits by randomly selecting one of the qubits incident on the missing bond to be measured in the $Z$ basis, 
The RHG lattice demonstrates significant robustness against bond failures. 
%, tolerating up to a maximum of $14.5\%$ bond failure rate with an adaptive measurement scheme in the absence of measurement errors. 
To handle missing bonds, two strategies can be employed: i) a non-adaptive scheme that requires no additional quantum processing and supports a maximum of $6.5\%$ bond failure rate, and ii) an adaptive scheme that measures qubits in an alternative basis to eliminate missing bonds and supports a maximum of $14.5\%$ bond failure rate~\cite{PhysRevA.97.030301}. The adaptive scheme works by associating bond failures with qubits, where one of the qubits incident on the missing bond is randomly selected to be measured in the Z basis, while the other is measured in the $X$ basis. The choice of which qubit to measure in $Z$ is random, and a qubit is measured in $Z$ only if the adjacent qubit has not been measured in $Z$ previously. 
This enhances the threshold compared to a non-adaptive approach where bond failures are managed by treating the qubits at each end of the failed bond as lost. The enhancement does however incur additional costs. Whereas all additional processing in the non-adaptive approach is handled classically during decoding without needing extra quantum processing, the adaptive scheme requires additional quantum processing, specifically the ability to alter the measurement basis during computations. It is important to note that the maximum bond failure tolerance thresholds mentioned above for the adaptive and non-adaptive measurement schemes are associated with the absence of measurement errors in FT-MQBC.

% For finite measurement errors, as the rate of measurement error rate increases, the bond failure tolerance threshold decreases. 
% This improves the threshold over a non-adaptive scheme that involves \kaushik{please complete...}. The improved threshold achieved with the adaptive scheme comes at the expense of requiring more quantum processing, specifically the capability to change the measurement basis during the computation.

\subsubsection{Noise Tolerance} In a scenario free from qubit noise errors, all the check operators of the kind in Eq.~(\ref{RHG_unit_stabilizer}) associated with the lattice exhibit a parity of +1. A single $Z$ error before and during measurement on a face qubit flips the parity of its unit cell's check operator $K_{\partial q}$ from $+1$ to $-1$, impacting both the affected unit cell and its neighbor. Notably, errors on two face qubits within the same unit cell do not alter its check operator parity but are detected by adjacent unit cells. This characteristic allows check operators to pinpoint the endpoints of error strings in the RHG lattice uniquely. 
A decoder then attempts to pair these endpoints to determine a correction strategy that minimizes the likelihood of logical errors. When there are no failed bonds, it leads to a measurement error threshold of approximately 2.9 percent ~\cite{PhysRevA.97.030301}. As the rate of bond failures increases, the measurement error threshold decreases for both adaptive and non-adaptive measurement schemes.  
% \kaushik{both approaches? Which two? Please clarify as I might be missing something.}.

%\kaushik{can we write this in terms of foliation?}

%Following measurements in the $X$ basis, the error syndrome is obtained by aggregating the outcomes of all check operators. This syndrome reveals the locations of error string endpoints on both primal and dual lattices. The decoder then attempts to pair these endpoints to determine a correction strategy that minimizes the likelihood of logical errors.

%  Fig.~\ref{fig:schematic} represents a unit cell of Raussendorf lattice. It has has nodes at each
% edge and face of a cube which are colored green and black respectively.   The lattice has two different kinds of layers,
% marked by green and black nodes. Reg bonds connect black nodes, and orange bonds connect green
% nodes. Interlayer bonds are colored blue.
% Fig. \ref{fig:schematic}, depicts a unit cell of the Raussendorf lattice. This cell contains nodes at each edge and face of a cube, represented as green and black colors, respectively. Two different kinds of layers exist within the lattice, marked by green and black nodes. Black nodes are connected by red bonds, while green nodes are linked by orange bonds. Blue bond connects the inter-layer bonds.

%\subsection{Two Layer Realization and Code Cycle}
\begin{figure}%[H]
       \centering
 \includegraphics[width=9cm]{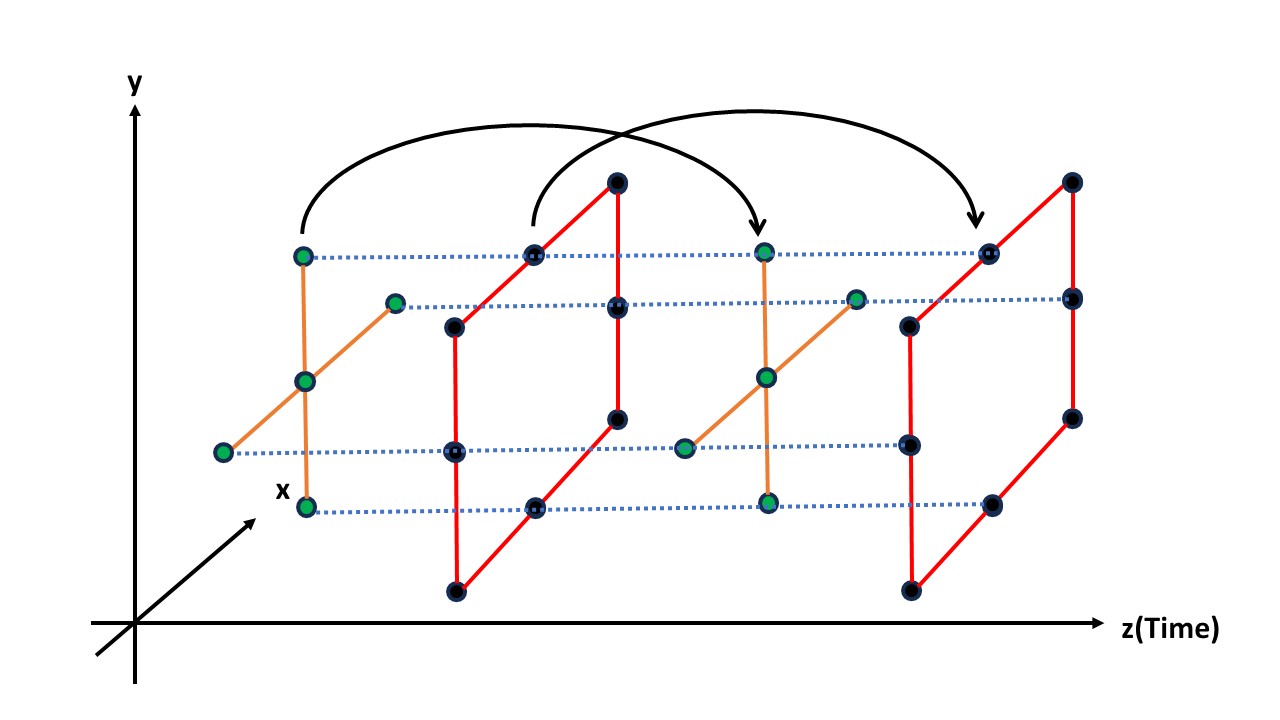}
        \caption{An adaptation of the RHG lattice into a 2+1D architecture. The lattice can be implemented using just two layers of 2D lattices, with memories being efficiently recycled post-measurement.}
        \label{fig:schematic1}
    \end{figure}

\subsection{Bi-Layered Realization\label{two layer}}
In MBQC, %the computational procedure involves sequentially measuring layers of the cluster state. The measurement of a layer leads to the teleporation of the processed quantum information onto the subsequent layer. 
since information propagates only one layer at a time, turns out it is sufficient to have at any given point of time only two layers prepared and connected~\cite{PhysRevLett.102.170501}. Subsequent to measuring one of the layers, the measured qubits can be reinitialized and reconnected to the other layer before measuring the other layer. The two can be perpetually measured, re-prepared and reconnected to execute arbitrarily long computations.
Such a two layer realization of the RHG lattice cluster state is depicted in Fig.~\ref{fig:schematic1}. To begin with, the two layers are created and connected, followed by measurement of the green-site layer. As a result, quantum information gets encoded into the black sites. The cycle continues with the re-connection of the green-site layer to the black sites through new blue bonds. The final step is measuring the values of the black sites, which completes \textit{\textbf{one code cycle}} (See Ref.\cite{PhysRevX.4.031022} and Supplementary Information of Ref.~\cite{Choi2019}).

\section{FT-MBQC over Optically Networked Trapped Ion Modules: Bond Failure Tolerance}\label{generation}
Networked architectures for FT-MBQC have been proposed and analyzed for various platforms such as NV centres, photonic qubits, ion-trap systems ~\cite{benjamin2009prospects,10160024,PhysRevA.89.022317}. The MUSIQC architecture \cite{PhysRevA.89.022317} pioneered its realization in optically networked trapped ion ELUs. In this work, we adapt the MUSIQC architecture to implement the RHG lattice using only two layers of ELUs, as discussed in Sec.~\ref{two layer}. We employ spatial and temporal multiplexing to generate the required remote Bell pairs, ensuring a sufficient bond success probability for fault tolerance. We then outline our methodology for constructing the RHG state and calculate the time needed to complete a code cycle in MBQC. Additionally, we examine a specific instance of a $2\times2\times1$ RHG lattice and estimate the resource requirements for its generation. To address the challenge of large distances between lattice sites, we incorporate quantum repeaters and perform a resource analysis with and without repeaters for this scenario. Finally, we evaluate the advantages of using quantum repeaters to efficiently handle large distances.

\subsection{Spatial and Time Multiplexed Remote Entanglement Generation}

%The rate of entanglement distribution through a pure loss bosonic channel with transmissivity $\eta$, a source repetition rate of $\frac{1}{\tau}$, and spatial multiplexing $M \in \mathbb{Z}^+$ is constrained by the entanglement distribution capacity~\cite{Pirandola2017}
% \begin{align}
% C_{\text{direct}}(\eta, M, \tau) = -\frac{M}{\tau} \log_2(1 - \eta) \ \text{ebits/s}.
% \end{align}
%To overcome this limit, various approaches have been developed, and a key distinction of the approaches is based on quantum repeaters~\cite{Muralidharan2016,PhysRevX.10.021071}.
Remote Bell pair generation between ELUs in the MUSIQC architecture is achieved via ion-photon entangled pair generation with communication ions, and photonic Bell state measurement (BSM) on the photons. The probability of successful remote Bell pair generation between a pair of ELUs separated by a distance $L$(km) is given by~\cite{vanLeent2022,labaymora2023reducing},
\begin{equation}
    p = \frac{1}{2}\eta_{\textrm{cc}}\eta_{\textrm{trans}}\eta_{\textrm{det}},
\end{equation}
where $\eta_{\textrm{cc}}$ denotes collection and coupling efficiency, $\eta_{\textrm{trans}}=10^{-(\alpha_{att}/10) L (\textrm{km})},\ \alpha_{att} = 0.2$  dB/km stands for the transmission efficiency, $\eta_{\textrm{det}}$ for detection efficiency, and the factor of $1/2$ corresponds to the success probability of a simple linear optical implementation of BSM (based on a beamsplitter and photodetectors). 
%The probability \( q \) for a successful entanglement swap between memory qubits in the same ELU is one, i.e., entanglement swapping can be realized deterministically, and also faithfully with the help of high-fidelity entangling gates and measurements designed for trapped ion qubits~\cite{Santra_2019}. 
%In this work, we incorporate both spatial and temporal multiplexing to enhance the probability of generating Bell pairs between two ELUs that are widely separated. 
This probability can be improved by multiplexing the entanglement generation attempts. Spatial multiplexing involves engaging multiple ions for ion-photon entanglement generation to support multiple simultaneous attempts, while time multiplexing involves repetition of attempts using same or different ions over blocks of time steps. The success probability of at least one successful remote Bell pair generation after $m\times M$ multiplexed attempts is given by~\cite{PhysRevA.105.022623},
\begin{equation}
    p'(m,M)={\left(1-(1-p)^{m M}\right)},
\end{equation}
where $m$ and $M$ are degrees of time and spatial multiplexing, respectively. 
It is essential to recognize that any improvement in the success probability due to multiplexing comes with a cost. An increase in spatial multiplexing comes with an increased budget of communication ion resources. An increase in time multiplexing degree \(m\) comes with a longer effective time step \(m\tau\) from \(\tau\), which may sometimes imply a smaller effective success rate over time.
 
%  $m = \frac{\tau_{c}}{\tau}$ where $\tau_{c}$ is the coherence time and $\tau$ is clock cycle duration, it denotes the rate
% at which the repeater nodes tries to generate ion-photon entanglement. 

% $q=1?$ is success probability of Bell swap operation which happens at a distance $L/2$?. So,

%  \begin{equation}
%     p(L, n, m)={\left(1-(1-p)^{m M}\right)}
% \end{equation}

\begin{figure}%[H]
       \centering
\includegraphics[width=8cm]{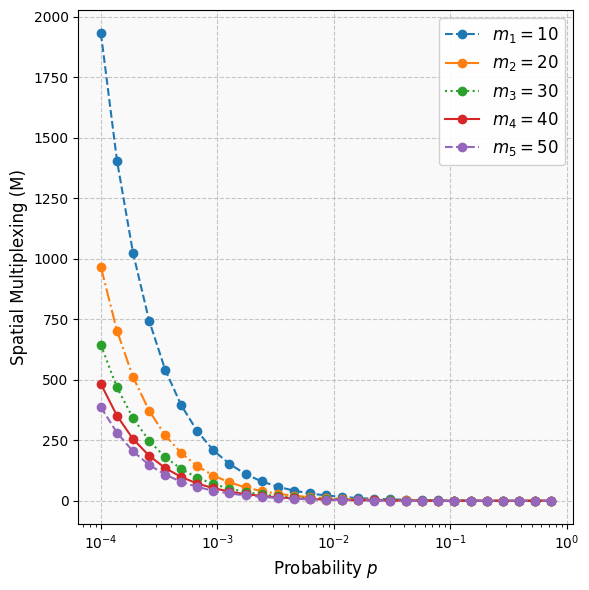}
        \caption{The plot shows the spatial multiplexing ($M$) vs probability of successful remote Bell pair generation ($p$) for various time multiplexing parameters $m_{i}$ where $i=1-5$ in order to attain $p_{th}=0.855$.}
         \label{p1}
    \end{figure}
    
%Paraphrase
 
Assuming the absence of measurement errors, the threshold of interest for the bond success rate for the RHG lattice denoted by $p_{\textrm{th}}$ is given by $1-0.145=0.855$ owing to the adaptive scheme of Ref.~\cite{PhysRevA.97.030301}. Given a success probability of a single remote entanglement generation attempt $p$, the product of spatial multiplexing parameter $M$ and time multiplexing parameter $m$ needed to attain $p_{\textrm{th}}$ is given by
% \begin{equation}
%     M = \dfrac{\log(1-p_{th})}{m(\log(1-0.5\times10^{-(\alpha_{att}/10) L}))}
% \end{equation}
\begin{equation}
    M m = \frac{\log(1 - p_{\text{th}})}{\left(\log\left(1 - p\right)\right)}.
    \label{Mm}
\end{equation}
Fig. \ref{p1} 
 shows the the spatial multiplexing that is required to attain $p_{th}=0.855$ as a function of $p$ for various time multiplexing parameters. It indicates that as the value of $p$ decreases, a higher degree of spatial multiplexing is necessary to meet the threshold, for any fixed time multiplexing. Additionally, it is possible to reach the same threshold with reduced spatial multiplexing, but this would require an increase in time multiplexing.

\subsection{Construction of the RHG lattice and Code Cycle Evaluation}\label{rhg_generation}

We now discuss the generation of the RHG lattice in optically networked trapped-ion ELUs aided by spatial and time multiplexing. We first describe the procedure of generating two layers of a unit cell, or, in other words, half a unit cell of the RHG lattice along the lines of the procedure used in Refs. \cite{PhysRevA.89.022317,RAUSSENDORF20062242}.

% \section{Resource Analysis}

%\subsection{Two layers}
%Note: 

\begin{figure}%[H]
       \centering
 \includegraphics[width=1.2\columnwidth]{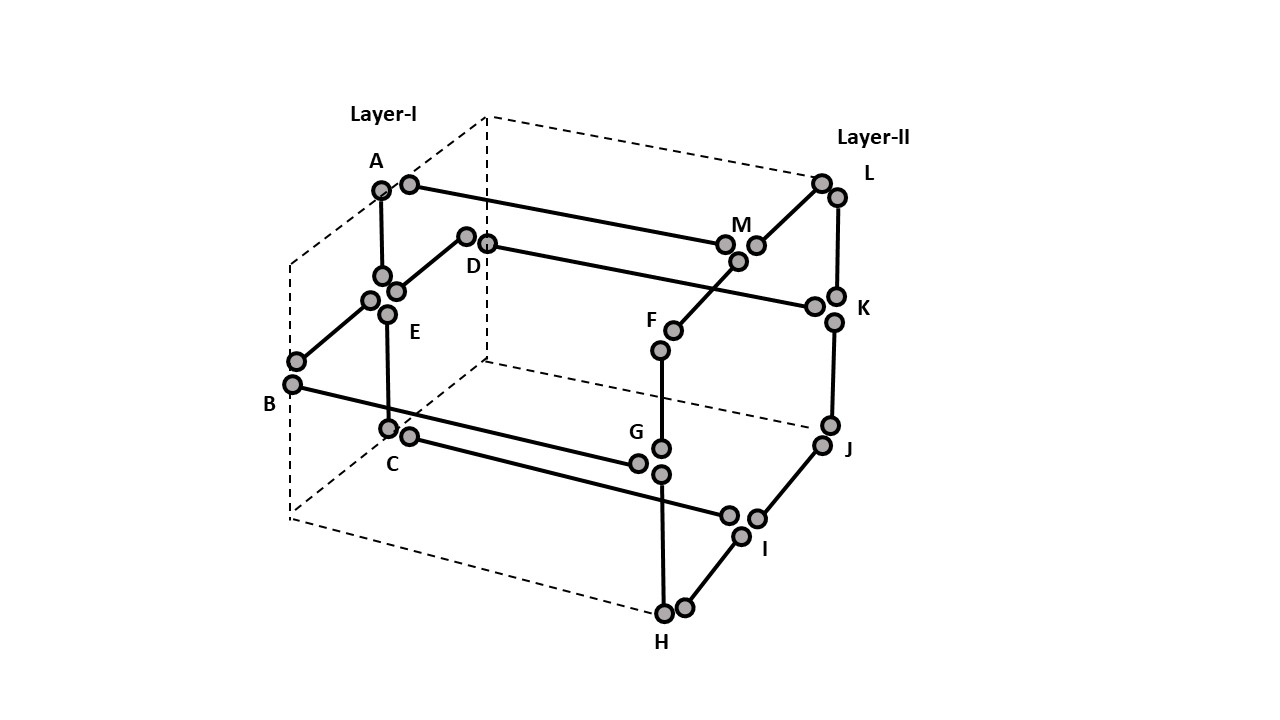}
        \caption{ $(A-E)$ are sites corresponding to Layer-I and $(F-M)$ are sites corresponding to Layer-II, each site correspond to one ELU. Two circles connected by a solid line represent a Bell pair between two layers.}
        \label{fig:schematic}
    \end{figure}

\begin{figure}%[H]
       \centering
 \includegraphics[width=\columnwidth]{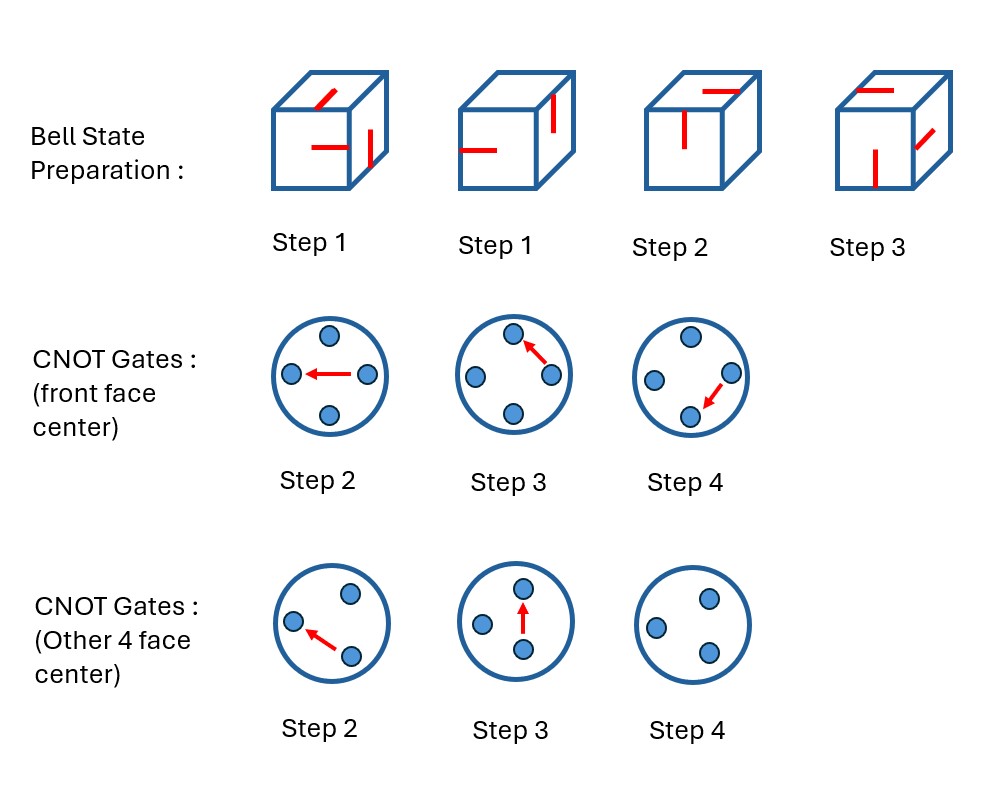}
        \caption{i) Steps to create RHG lattice. }
        \label{fig:schematic1_1}
    \end{figure}
% Once, we have successfully generated Bell pairs between all the nodes as described in Fig.~\ref{fig:schematic}. We adapt a similar approach of Ref.\cite{PhysRevA.89.022317} based on Ref.\cite{RAUSSENDORF20062242} but generate only two layers of RHG state.
% At each node, we perform CNOT operations and X/Z measurements depending on whether the node is a face qubit or an edge qubit as elaborately detailed in Ref.\cite{PhysRevA.89.022317}. The only difference being, we would need less number of CNOTs at some of the nodes, as we prepare only two layers as compared to three layers. We can repeat these two layers in time to perform computation by reusing the resources. With the above modifications, we estimate the resource requirements for generating the architecture and calculate the the duration of one code cycle. 

\begin{figure}%[H]
       \centering
 \includegraphics[width=0.90\columnwidth]{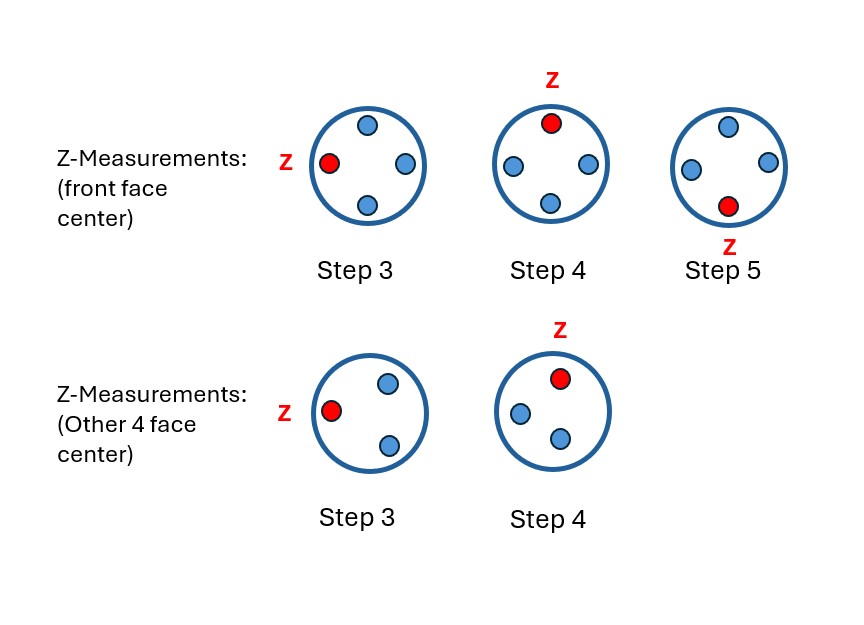}
        \caption{ii) Steps to create RHG lattice.%\kaushik{Shouldn't the other 4 faces also be measured in $Z$ basis? It is the edge centers that need to be measured in $X$ basis. For that matter, you haven't not shown edge center steps. Do you think we should show? Or can we mention in the text that we show just face centers, whereas edge centers are similar except direction of cnot gates with a common control vs a common target and X instead of Z measurement? Also, Fig 8 and 9 can be combined into one figure?}}
        }\label{fig:schematic1_2}
    \end{figure}

Below is a brief overview of the tasks involved in creating two layers of a unit cell of RHG state (see Fig. \ref{fig:schematic}),
\begin{itemize}
    \item  Generation of Bell pairs between each pair of adjacent lattice sites within and across layers I and II (all sites assumed to be equidistant) using spatial and time multiplexing.
    
    \item Performing CNOT gates operations and $X/Z$ measurements at each site, taking into account the entanglement links with adjacent sites.
\end{itemize}

The procedure we consider is an adaptation of the procedure used in Ref.~\cite{PhysRevA.89.022317} for the generation of the RHG state, and is as shown in Figs.~\ref{fig:schematic1_1} and \ref{fig:schematic1_2}. Here, the generation of Bell pairs between nodes is spread across three time steps. This is followed by CNOT gates and local measurements in subsequent time steps. All operations are scheduled such that each ancilla qubit (i.e., all except one qubit per node) in an ELU is active for exactly three time steps, namely, they are initialized (as part of a remote Bell pair) in one time step, undergo CNOT operation in another time step, and get measured off in another time step. Throughout this protocol, no qubit is ever left idle in the ELUs. The figures (~\ref{fig:schematic1_1} and \ref{fig:schematic1_2}) present a front view featuring three faces. %Note that the other three faces of the unit cell will actually be accounted for as part of neighboring unit cells, namely the cells below, to the left and behind.
   % (See Figs.~\ref{fig:schematic1_1} and \ref{fig:schematic1_2}). This completes the preparation of the two layers (a half unit cell) of the RHG state as shown in Fig.~\ref{fig:schematic}.
    
    % \textit{ Notably, fewer CNOTs are required at some nodes due to the preparation of only two layers instead of three.}

    % \item Estimation of resource requirements for generating the architecture with the mentioned modifications.
    
    % \item Calculation of the duration of one code cycle.
%\end{itemize}

In Figs.~\ref{fig:schematic1_1} and \ref{fig:schematic1_2}, we illustrate the procedure at the face center nodes of the RHG lattice. These are represented in Fig.~\ref{fig:schematic} as node $E$ (front face node), and nodes $G$, $M$, $I$, and $K$ (other face-center nodes). When adapting the three-dimensional RHG lattice into two layers (as discussed in Ref.~\cite{PhysRevX.4.031022}), the number of nearest neighbors for face-center qubits becomes either 3 or 4, depending on their position. Note that the procedure at the edge center nodes (not shown here) is similar, but with a few points of distinction. 
%It involves CNOT gates, with a point of difference being that all CNOT gates at an edge center node will have a common target as opposed to a common control as in the case of face center nodes, and $X$ measurements in place of $Z$ measurements.
The key distinction lies in how the CNOT gates operate. Whereas at face-center nodes, one qubit serves as the control for all CNOTs within the ELUs, and the remaining ancilla qubits are measured in the $Z$ basis, at edge center nodes, a qubit within the ELU acts as the target of all CNOT gates, with the remaining qubits being measured in the $X$ basis.%This procedure follows the approach outlined in Ref.~\cite{PhysRevA.89.022317}.

 FT-MBQC over the bi-layered realization of the RHG lattice then proceeds by first measuring the qubits in Layer-I. Following this, new Bell pairs are once again remotely generated between Layer-II and Layer-I by photonic mediation, and the necessary CNOT gates and Pauli measurements are applied to generate the next half unit cell. Then, the qubits in Layer-II are measured. The process is iteratively repeated to generate as many layers as needed for the computation being executed by reusing resources over time across the two layers of trapped ion ELUs. Evidently, the two layers end up dividing between them the even and odd computational steps, respectively. The approach thus allows for execution of arbitrarily long computations with finite number of physical qubits at the trapped-ion ELUs~\cite{PhysRevLett.102.170501,PhysRevX.4.031022}.

% For our calculations, we consider only a unit cell, however it can be generalized for arbitrary finite dimension RHG lattice.

 % and reuse the resources to generate further layers in time. 

%We consider first two layers of RHG lattice, where each node are equidistant from one another. 

% In this work, we propose a method to build the RHG state in an ion-trap system.
Before we proceed to analyze the code cycle duration and resource requirements, we introduce the various relevant timing parameters of the system, and their typical values, tabulated in Tables I and II, respectively. %We denote the clock-cycle duration by $\tau$, which is the rate at which communication ions at the ELUs attempt ion-photon entangled pair generation. 
%The gate operation and measurement takes time may vary across experimental platforms. We denote the coherence time of the communication and memory ions are denoted by $\tau_{c}$ and $\tau_{m}$ respectively. The various parameters used in this work and their associated meaning are shown in the table below. 
\begin{center}
    \begin{tabular}{cl} 
        \hline
        Timing Parameter & Associated Meaning \\
        \hline
        $\tau$ & Clock cycle duration
        \\
        $\tau_m$ & Memory ion lifetime \\
         $\tau_H$ & Heralding Time \\
        $\tau_c$ & Communication ion lifetime \\
        $\tau_a$ & Operation time for single-qubit gate \\
        $\tau_b$ & Operation time for two-qubit gate \\
        $\tau_{d}$ & Qubit Measurement Time \\
         $n$ & Refractive Index of medium \\
        
        \hline
    \end{tabular}
\end{center}

%\vspace{6pt}  % add vertical space for separation

\textbf{Table I.} \label{table_1}Time parameters associated with quantum operations in trapped-ion system.

\begin{center}
    \begin{tabular}{cl} 
        \hline
        Timing Parameter & Associated Value \\
        \hline
        $\tau$ & 10 $\mu s$
        \\
        $\tau_a$ & 1 $\mu s$ \\
        $\tau_b$ & 10 $\mu s$ \\
        $\tau_{d}$ & 30 $\mu s$ \\
         $n$ & 1.5 \\
        
        \hline
    \end{tabular}
\end{center}

%\vspace{6pt}  % add vertical space for separation

\textbf{Table II.} Typical values  of time  parameters associated with quantum operations in trapped-ion system~\cite{PhysRevA.89.022317}.
\vspace{6pt}

% The RHG lattice cell in Fig.~\ref{fig:schematic} has $5$ sites in the first layer and $8$ sites in the second layer. 

In order to generate an entanglement pair between two sites, we attempt spatially multiplexed Bell pairs generation during every clock cycle of duration $\tau$ for $m$ clock cycles. We know that the Bell pair generation succeeds only probabilistically, but can be heralded. 
%After at least one successful generation of Bell pair between two sites, we need to evaluate the time to know if the generation is successful. 
If the distance between two sites is $L$, the heralding time $\tau_{H}$ is given by,
\begin{equation}
    \tau_{H} = \dfrac{L}{2c/n} + \dfrac{L}{2c/n} = \dfrac{nL}{c},
\end{equation} 
where $L/2$ is the distance between each ELU and the virtual node between a pair of ELUs, where Bell state measurement is performed and $c/n$ is the speed of light in an optical fiber, where $n$ is the refractive index of the medium. We store the entanglement in memory ions, which requires performing a SWAP gate~\cite{Santra_2019}. This gate can be decomposed into three CNOT gates, taking a total time of \( 3\tau_{b} \).
Evidently, the maximum time $T$ required for the generation of a Bell pair between two sites at the RHG lattice is given by,
\begin{equation}
    T = m\tau + \tau_{H} + 3\tau_{b}.
    \label{timestep}
\end{equation}
It is intrinsically assumed that the decoherence time of communication ions $\tau_{c}$ is greater than $T$. For the typical values given in Table II, we note that each of the steps (Step 1 to Step 5), some of which involve remote Bell pair generation alone, while some involve Bell pair generation and CNOT gates, and some involve CNOT gates and local measurements alone, can be completed in a time duration not exceeding \( T \). That is, it can be seen that,
\[
T \geq \max \{ 
 T_{1}, T_{2}, T_{3}, T_{4}, T_{5} \},
\]
where \( T_{i} \) represents the time to complete the \( i^{th} \) step, with \( i \in \{1,\ldots,5\} \).% because $T$ involves multiple operations including gate operations and Bell pair generation.

 % As described earlier for generating the two layers, we apply CNOT gates and perform $X/Z$ measurements to accomplish STEP 2. This require performing a maximum of three CNOT gates at the face center of first layer which would require a time of $3\tau_b$ since gates need to be performed sequentially as they share the same control qubit and three single qubit measurements ($\tau_d$) at the face qubits. Measurements need to be performed after the operation of CNOT gates. So, the total time $\tau_{2}$ (STEP 2 completion) is given by

 According to the definition of code cycle, we first generate the two layers as shown in Fig.~\ref{fig:schematic} which requires a total of $5$ time steps of duration $T$, 
\begin{equation}
    \tau_{1} = 5T = 5(m\tau + \tau_{H} + 3\tau_{b}) 
\end{equation}
Once the two layers are generated, the computation proceeds by measuring the first layer of the cluster, which results in the teleportation of information to the second layer. The total time $\tau_{2}$ until the completion of the first layer measurement is given by,
\begin{equation}
    \tau_{2} = \tau_{1} + \tau_{d} = 5(m\tau + \tau_{H} + 3\tau_{b}) + \tau_{d}
\end{equation}
%After the measurements, the physical qubits that were measured can be reconnected to the second layer, leading to the teleportation of information to next layer. 
Once the first layer is measured off, it needs to be reconnected to the second layer. 
%In this work, we consider both the cases when we can selectively measure the qubits and the other case when all qubits need to be measured off. 
Though multiple remote Bell pairs may have been successfully generated between the two layers previously of which only one got used in the cluster state, for simplicity of analysis, we discard the unused ones, and analyze the time duration with respect to fresh generation of remote Bell pairs for the re-connection of the two layers during the code cycle. The re-generation of the connections between layer II and layer I, can again be done within $5$ time steps of duration $T
$. The net time elapsed until the reconnection of the layers is thus given by,
% The time taken to generate these Bell pairs in $m\tau$ time steps and subsequently to perform the bond heralding measurements is given by
\begin{equation}
    \tau_{3} = \tau_{2} + 5T = 10(m\tau + \tau_{H} + 3\tau_{b}) + \tau_{d}.
\end{equation}
% The net time elapsed thus far is therefore given by
% \begin{equation}
%     \tau_{4} = \tau_{3} + \tau_{L1}^{\prime} = (2m)\tau + 2\tau_{H} + 12\tau_{b}+3\tau_{d} 
% \end{equation}
After Layer-II is connected to Layer-I, layer II is measured off, marking the completion of one code cycle. The net time elapsed until this is given by,
\begin{equation}
    \tau_{4} = \tau_{3} + \tau_{d} = 10(m\tau + \tau_{H} + 3\tau_{b}) + 2\tau_{d} \label{tau_5}. 
\end{equation}

By substituting
\begin{equation}
    m(M,p) = \frac{\log(1 - p_{\text{th}})}{M(\log\left(1 - p)\right)}
\end{equation}
in Eq.~(\ref{tau_5}), we get the code cycle in terms of the spatial multiplexing parameter,
% \begin{equation}
% \tau_{4} =   10((\frac{\log(1 - p_{\text{th}})}{M(\log\left(1 - p)\right)})\tau + \tau_{H} + 3\tau_{b}) + 2\tau_{d} 
% \end{equation}
\begin{equation}
\tau_{4} = 10 \left( \frac{\log(1 - p_{\text{th}})}{M \log(1 - p)} \tau + \tau_{H} + 3\tau_{b} \right) + 2\tau_{d}
\label{code_cyc}
\end{equation}
% \begin{equation}
%     \tau_{5} = (2\frac{\log(1 - p_{\text{th}})}{M\left(\log\left(1 - 0.5 \times 10^{-\tfrac{\alpha_{\text{att}}}{10} L}\right)\right)})\tau + 2\frac{nL}{c} + 12\tau_{b}+4\tau_{d}
% \end{equation}
% \begin{align}
% \tau_{5} &= \left(2\frac{\log(1 - p_{\text{th}})}{M(\log(1 - p)}\right)\tau \nonumber\\
% &+ 2\frac{nL}{c} + 12\tau_{b}+4\tau_{d}\label{tau_5_M}
% \end{align}
The case \( M = 1 \) here corresponds to a `repeat until success' strategy, where a single communication ion at each site attempts Bell pair generation repeatedly over time. In this approach, ion-photon entangled pair generation is attempted once every clock cycle (of duration \(\tau\) \(\mu\)s), over a block of \(m\) clock cycles, with only one communication ion being repeatedly used, as was the case in the MUSIQC architecture of Ref.~\cite{PhysRevA.89.022317}. $M>1$ corresponds to attempting Bell pair generation between pairs of sites multiple times in parallel during each clock cycle, clearly incurring the usage of multiple communication ions. Based on Eq.~(\ref{code_cyc}), an increase in $M$ should result in a shorter code cycle time. 
\begin{figure}%[H]
       \centering 
\includegraphics[width=\columnwidth]{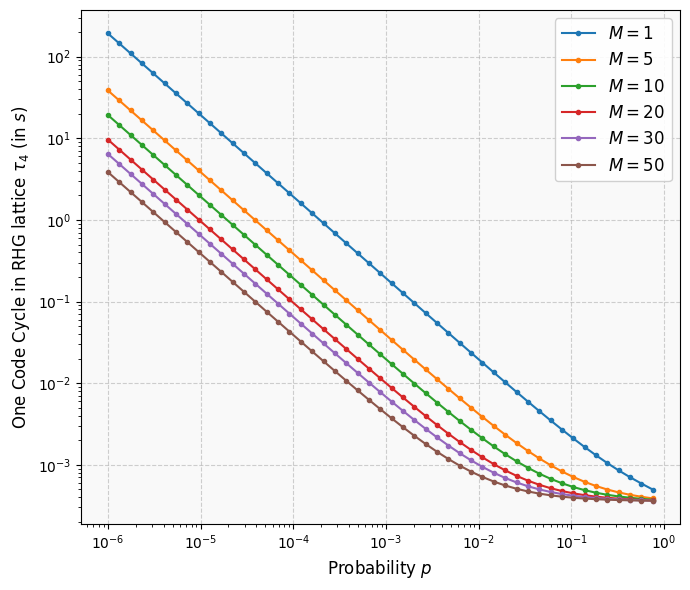}
        \caption{The plot shows time duration of one code cycle vs probability of successful remote Bell pair generation ($p$) for various spatial multiplexing parameters $M = \{1,5,10,20,30,50\}$  in order to attain $p_{th}=0.855$ for $L=1 
 \mu m$.}
         \label{p2}
    \end{figure}

\begin{figure}%[H]
       \centering 
\includegraphics[width=\columnwidth]{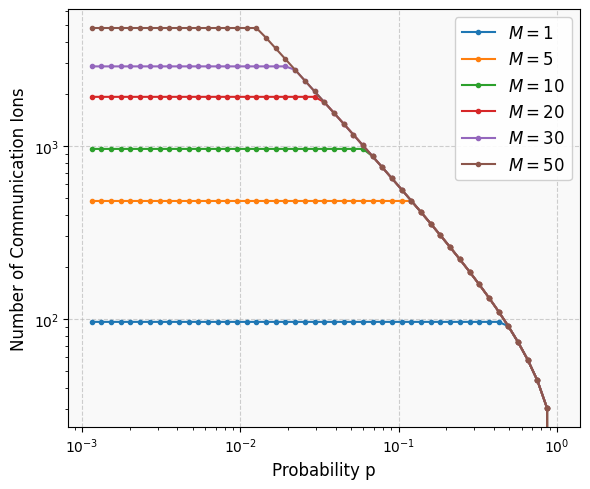}
        \caption{The plot shows number of communication ions required vs probability of successful remote Bell pair generation ($p$) for various spatial multiplexing parameters $M = \{1,5,10,20,30,50\}$  in order to attain $p_{th}=0.855$ for $L=1 
 \mu m$.}
         \label{ionsvsp}
    \end{figure}
%In Fig. \ref{p2},

%In the context of spatial multiplexing, using more communication ions surpasses the `repeat until success' scheme. Spatial multiplexing enables simultaneous transmission of multiple communication channels through distinct paths, allowing concurrent generation of entangled pairs between sites. This significantly reduces the time required for entanglement generation compared to the sequential attempts in the `repeat until success' scheme. Consequently, spatial multiplexing achieves faster cycle times for the code, leading to improved  performance in generating entangled pairs.

Figure~\ref{p2} illustrates the code cycle duration for the RHG lattice (shown in Fig. \ref{fig:schematic}) versus the probability of successful remote Bell pair generation between two sites for different values of spatial multiplexing parameter $M$. We observe that for a low success probability of Bell pair generation, the code cycle duration is very high, while for a high success probability of Bell pair generation, it is small. Additionally, higher multiplexing parameters result in a shorter code cycle compared to lower multiplexing parameters for the same Bell pair generation probability.

\begin{figure}%[H]
       \centering
      
\includegraphics[width=\columnwidth]{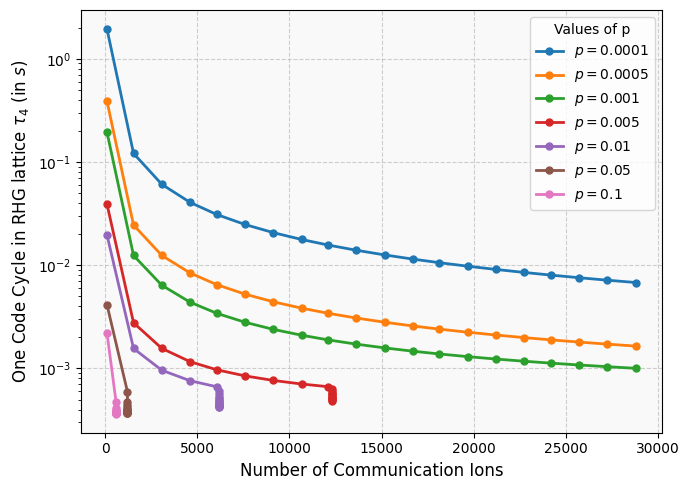}
 %        \caption{The plot shows time duration of one code cycle vs probability of successful remote Bell pair generation ($p$) for various spatial multiplexing parameters ($M$)  in order to attain $p_{th}=0.855$ for $L=1 
 % \mu m$.}
 \caption{The plot shows the relationship between one code cycle $\tau_4$ in the RHG lattice and the number of communication ions $32M \min(m, j)$  shown in Fig.~\ref{fig:schematic} for various spatial multiplexing parameters $M$ ($20$ evenly spaced integers from $1$ to $300$) and probabilities $p$ in order to attain $p_{th}=0.855$ for $L=1 
 \mu m$. The values of \( p \) range from \( 10^{-4} \) to \( 10^{-1} \).
}

         \label{p3}
    \end{figure}

Next, we discuss the resource requirements. We will assume a scenario, where memory ions are available in abundance at ELUs, but communication ions are scarce and will be reused as much as possible. Such a scenario is practically relevant, as communication ions in an ELU have to be far spaced out from one other to avoid resonant re-absorption of emitted photons. One possible mode of operation under this assumption is where the quantum information of every communication ion after it has been involved in remote Bell pair generation is immediately stored in a memory ion. In this case, firstly, the memory ion requirement associated with remote Bell pair generation between a pair of sites is given by $2Mm$ (i.e., $Mm$ at each site), since that is the total number of Bell pair generation attempts. Since Fig.~\ref{fig:schematic} involves Bell pair generation across 16 pairs of sites, the cumulative number of memory ions required across the ELUs is given by $32Mm$. Secondly, the communication ion requirement associated with remote Bell pair generation between a pair of sites is given by $2M\min(j,m)$, where $j=3\tau_b/\tau$ (which for the values from Table II equals 3). This is because, a batch of $M$ communication ions involved in a Bell pair generation attempt can be reused after $j$ clock cycles~\cite{PhysRevA.105.022623}. The cumulative number of communication ions required across ELUs, thus, is given by $32M\min(j,m)$. %\kaushik{This is actually only 4+12/2=10, because rest will be accounted for by neighboring cells on left or right, and top or bottom, so $32Mm\rightarrow 20Mm$ and 20 memory ions} 
%we would need $32Mm$ communication ions and $32$ memory ions to store the entanglement. 

% Figure~\ref{p3}, shows  the relationship between the number of communication ions $32Mm(M, p)$ and one code cycle in the RHG lattice $\tau_4$ across various spatial multiplexing parameters $M$. The $x$-axis in the figure represents one code cycle $\tau_4$ in seconds ($s$) on a logarithmic scale, while the $y$-axis represents the number of communication ions $32Mm(M, p)$ also on a logarithmic scale. We note that the probability $p$ decreases from left to right along the $x$-axis. The plot shows how the code cycle duration and the number of communication ions change as the success probability, \( p \), decreases. A lower \( p \) means more attempts are needed to generate at least one successful Bell pair, leading to longer code cycles and a higher number of communication ions required to compensate for the reduced success rate. Conversely, higher \( p \) values correspond to fewer attempts, shortening the code cycle and reducing the number of communication ions. Thus, the plot highlights how the resource demands for generating the RHG lattice depend on both the success probability \( p \) and the spatial multiplexing parameter \( M \).

Figure~\ref{ionsvsp} illustrates how the number of required communication ions varies with the remote Bell pair generation probability $p$ for different spatial multiplexing parameters \( M \). As $p$ increases, initially, the number of communication ions remains constant up to a certain threshold value, beyond which it decreases. This pattern arises due to the function \( \min(j, m) \) taking on the constant value \( j \) when \( m > j \), but otherwise taking on the time multiplexing parameter \( m \), which is determined by both the probability \( p \) and the spatial multiplexing parameter \( M \).

Figure~\ref{p3} illustrates the relationship between the code cycle duration \( \tau_4 \) and the number of communication ions in the RHG lattice for different values of the spatial multiplexing parameter \( M \). The $y$-axis represents the code cycle duration, \( \tau_4 \), in seconds ($s$), on a logarithmic scale, while the $x$-axis represents the number of communication ions \( 32M \min(j, m) \). We plot the relationship for different values of the remote Bell pair generation probability $p$. As \( p \) decreases, more attempts are required to generate at least one successful Bell pair, resulting in longer code cycles and a higher number of communication ions needed to compensate for the lower success rate. Conversely, higher values of \( p \) correspond to fewer required attempts, reducing both the code cycle duration and the number of communication ions. The plot demonstrates how the resource demands for generating the RHG lattice are influenced by both the success probability \( p \) and the spatial multiplexing parameter \( M \). We observe that the required number of communication ions becomes constant beyond a certain value of the spatial multiplexing parameter \( M \). This is because \( \min(j, m) \) takes the value \( m \) when \( j > m \), resulting in the quantity (i.e., the required number of communication ions) becoming the product of spatial and temporal multiplexing parameters \( M m \), which is a constant for any fixed value of probability \( p \) (see Eq.~\ref{Mm}). Consequently, any increase in \( M \) is compensated by a decrease in \( m \), which results in a decrease in the code cycle duration, while preserving the constant value of the product \( M m \).

\subsection{Repeater Enhanced Scheme}\label{repeater}

When the lattice sites (ELUs) are separated by large distances, the resource requirement to support a high $p_{th}$ increases drastically. In order to alleviate this issue, we can place multiple quantum repeaters in between pairs of adjacent sites.
\begin{figure}[H]
       \centering
      
\includegraphics[width=9cm]{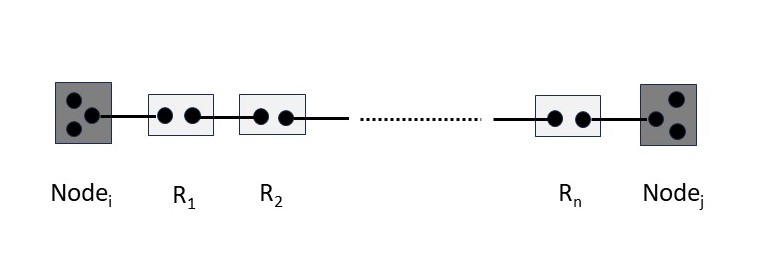}
        \caption{Between two sites ($i$ and $j$), we place $n$ repeaters labelled as $R_{1}$ to $R_{n}$}
         \label{p4}
    \end{figure}

For an $n$ repeater chain between two trapped-ion ELUs, the distance between two adjacent repeaters would be $L/(n+1)$, and the probability of successful generation of a remote Bell pair between the repeaters is given by~\cite{PhysRevA.105.022623},
\begin{equation}
   p_{0}' = \frac{1}{2} \cdot 10^{-0.02 \tfrac{L}{n+1}} = \frac{1}{2} \left(2p_{0}\right)^{\frac{1}{n+1}}
\end{equation}
where $p_{0} = \frac{1}{2} \cdot 10^{-0.02L}$. 
With $M^{\prime}$ spatial multiplexing and $m$ time multiplexing, assuming deterministic entanglement swap operations at the repeater sites, the success probability of generating at least one successful Bell pair between the end modules is given by,
\begin{equation}
    p(m, M', p_{0}') = \left[1 - \left(1 - p_{0}'\right)^{mM'}\right]^{n+1}.
\end{equation}
For $
    p(m,M',p_{o}') = p_{th}$, the spatial multiplexing $M^{\prime}$
 is given by,
\begin{equation}
    M' = \frac{\log\left(1 - p_{\text{th}}^{\tfrac{1}{n+1}}\right)}{m \log\left(1 - p_{0}'\right)}.
\end{equation}

To store all the attempted entanglement pairs between two ELUs using Bell pair generation with time multiplexing $m$, \( Mm \) memory ions are needed in the case without quantum repeaters, \( M^{\prime}m \) are needed in the case with $n$ quantum repeaters, given by,
\begin{align}
    Mm &=  \frac{\log(1 - p_{\text{th}})}{\log\left(1 - p_{0}\right)}, \\
    M^{\prime}m &= \frac{\log\left(1 - p_{\text{th}}^{\tfrac{1}{n+1}}\right)}{\log\left(1 - p_{0}'\right)}.
\end{align}
% \begin{figure}[H]
%        \centering
      
\begin{figure}%[H]
       \centering 
\includegraphics[width=\columnwidth]{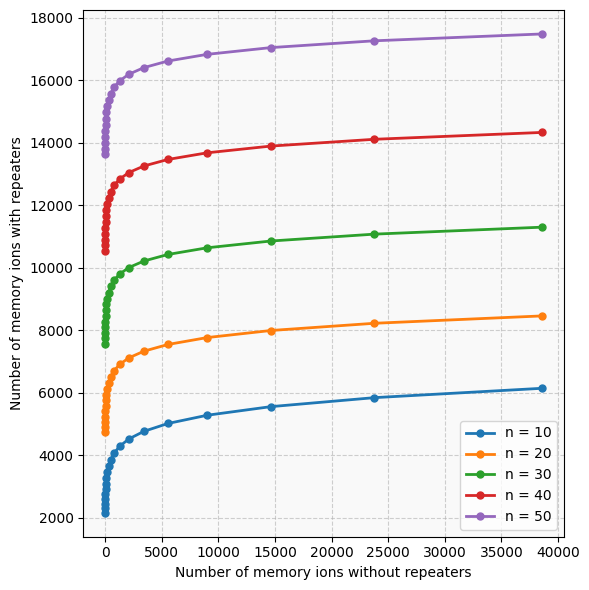}
        \caption{The plot shows the cumulative number of memory ions required for 2 layers of RHG lattice shown in Fig.~\ref{fig:schematic} where $y$ axis is with $n$ quantum repeaters between two sites and $x$ axis is without repeaters.  Each point from left to right corresponds a distance from 20 evenly spaced values ranging from 0 to 200 km in order to support $p_{th}=0.855$ for different values of number of repeaters ($n$) placed between two sites.}
         \label{ratio}
    \end{figure}

% \includegraphics[width=8cm]{ratio.jpg}
%         \caption{The plot shows the the number of communication ions required at the nodes of RHG unit cell for generating one Bell pair in $y$ axis with repeaters vs without repeaters in $x$ axis which each point from left to right corresponds a distance from 20 evenly spaced values ranging from 0 to 200 km in order to support $p_{th}=0.855$ for different values of number of repeaters $(n)$ placed between two nodes.}
%          \label{ratio1}
%     \end{figure}
% Analyzing the plot in Fig.~\ref{ratio}, we note an increased count of communication ions at shorter distances. In contrast, for larger distances, there is a substantial reduction in resource requirements, attributable to the incorporation of quantum repeaters.
With $n$ quantum repeaters between two end modules, we would need to generate $(n+1)$ Bell pairs. Thus, we would be needing $16\times2(n+1)M'm$ memory ions cumulatively across the $n$ quantum repeaters to store all the attempted entanglement pairs towards generating the $16$ Bell pairs required for two layers of the RHG lattice shown in Fig.~\ref{fig:schematic}. We denote this number as
\begin{equation}
    N^{\textrm{mem}}_{R} = 32(n+1) \frac{\log\left(1 - p_{\text{th}}^{\tfrac{1}{n+1}}\right)}{\log\left(1 - p_{0}'\right)}.
\end{equation}
%where $N_{R}$ is the cumulative number of memory ions across $n$ quantum repeaters between two end modules.
On the other hand, without quantum repeaters between the end modules, the number of memory ions required %for generating $16$ Bell pairs for the RHG lattice state in Fig.~\ref{fig:schematic} 
is $32Mm$, i.e.,
\begin{equation}
     N^{\textrm{mem}}_{WR} = 32 \cdot \frac{\log(1 - p_{\text{th}})}{\log(1 - p_{0})}.
\end{equation}
Fig.~\ref{ratio} shows the number of memory ions required for $2$ layers of RHG lattice in Fig.~\ref{fig:schematic} with and without $(n)$ quantum repeaters between pairs of adjacent sites. We observe that although quantum repeaters require more memory ions at shorter distances, they drastically reduce resource requirements over longer distances.
%where $N_{WR}$ is the total number of memory ions without quantum repeaters between the two modules (ELUs).
%After successfully establishing connections between two end modules (ELUs) using Bell pairs, we proceed to transfer them to memory ions due to their longer coherence times. Consequently, to generate this particular RHG lattice, comprising of 16 Bell pairs, a total of 32 memory ions will be required.

\begin{figure}%[H]
       \centering
      
\includegraphics[width=\columnwidth]{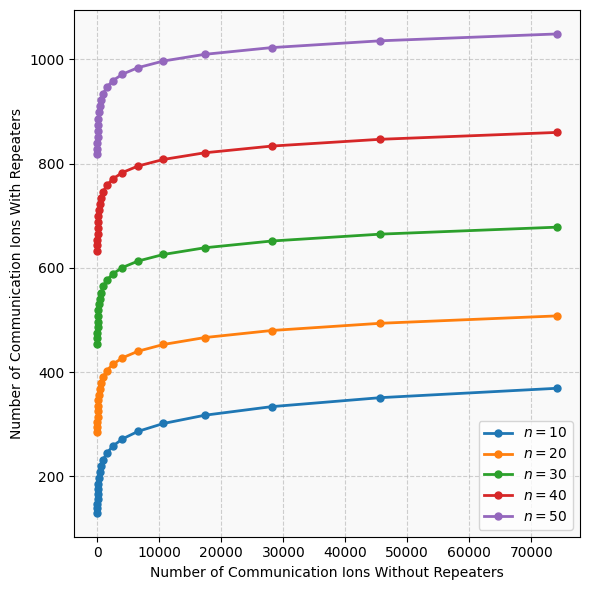}
        \caption{The plot shows the cumulative number of communication ions required for 2 layers of RHG lattice shown in Fig.~\ref{fig:schematic} where $y$ axis is with $n$ quantum repeaters between two sites and $x$ axis is without repeaters for temporal multiplexing parameter $m=50$.  Each point from left to right corresponds a distance from 20 evenly spaced values ranging from 0 to 200 km in order to support $p_{th}=0.855$ for different values of number of repeaters ($n$) placed between two sites.}
         \label{ratio1}
    \end{figure}

We also note that in the case with repeaters, the cumulative communication ion count across the repeater chain would amount to $N^{\textrm{comm}}_{R}=32(n+1)M' \min(j,m)$ as opposed to $N^{\textrm{comm}}_{WR}=32M \min(j, m)$ in the case without repeaters, where $j=3\tau_b/\tau$, which for the typical values of time parameters from Table II is 3. 
Fig.~\ref{ratio1} shows the number of communication ions required for two layers of the RHG lattice in Fig.~\ref{fig:schematic}, with and without \( n \) quantum repeaters between adjacent sites. We observe that, while quantum repeaters require fewer communication ions at all distances, they dramatically reduce resource requirements over longer distances. This highlights the  advantage of quantum repeaters in minimizing resource consumption for long-distance distributed quantum computing.

% \section{Extension to arbitary dimension Raussendorf Lattice}
% It is evident that placing quantum repeaters drastically reduces the resource requirements.

\section{FT-MBQC over Optically Networked Trapped Ion Modules: Noise Tolerance}\label{fault}

% In the previous section, we dealt with a strategy to deal with loss errors. However, for achieving fault tolerance in real scenarios, we need to deal both with computation errors. We know that there exits high thresholds in Fault-Tolerant Quantum Computation (FTQC) for both computational errors and loss errors which serves as a crucial benchmark for practical implementations and is attributed to the substantial redundancy within the surface codes that form the basis of Raussendorf's FTQC scheme.
In the previous section, we addressed mitigating bond failures in the RHG lattice using multiplexed remote Bell pair generation. To attain fault tolerance in realistic conditions, in addition to bond failure errors, it is imperative to also address computational errors. FT-MBQC imposes a stringent 2-dimensional threshold region for computational errors and bond failure errors, serving as a pivotal benchmark for practical implementations~\cite{PhysRevLett.105.200502,PhysRevA.97.030301}. These thresholds are indicative of the significant redundancy embedded within the three-dimensional error correcting codes that fundamentally constitute the framework of the FTQC scheme proposed by Raussendorf et al.~\cite{PhysRevA.71.062313, RAUSSENDORF20062242,Raussendorf_2007}.

% In this section, we deal with computational errors that occurs in preparation of Bell pairs, measurement and gate operation on individual qubits and use the inequality on error probability
% of the stabilizer measurement process for the 3D cluster state to probe the depolarising noise while considering preparation of imperfect Bell pair considering which may arise due to hardware and channel imperfections when dealing with
% practical entanglement swapping links.

In this section, we consider the noise errors that occur in the preparation of Bell pairs, gate operations and qubit measurements, and decoherence with time, and derive  an inequality governing the error probability of the stabilizer measurement process for the 3D cluster state, specifically tailored to our approach. In contrast to Ref.~\cite{PhysRevA.89.022317}, which considers a different noise model, our focus is especially on the depolarizing noise that arises from a realistic photon-mediated remote entanglement generation process. The imperfect Bell states thus produced account for practical limitations in the optical channels and detection hardware~\cite{PhysRevResearch.5.033149}.
%the practical hardware limitations and their transmission through imperfect optical channels~\cite{PhysRevResearch.5.033149}.

\subsection{Remote Entanglement Generation}
% For Bell state preparation, entanglement swap is required in order to manipulate the quantum states of photons in a way that allows the entanglement of two separate photon pairs. In this process, the photonic qubits undergo a mixing process on a balanced ($50:50$) beamsplitter, which serves to erase the information regarding their source or path. Subsequently, they are subjected to detection through photon number-resolving detectors. The identification of particular click patterns in the detectors signifies the successful generation of an entangled state. This entangled state is shared among quantum memories, and the associated information about the parity of the entangled state is obtained through this detection process.

For the preparation of remote ion-ion Bell pairs via generation of ion-photon Bell pairs, photonic entanglement swapping is essential~\cite{PhysRevA.72.052330,PhysRevLett.91.110405}. This process allows for the manipulation of the quantum states of photons, enabling the entanglement of two remote ions. In this process, the photonic qubits undergo a mixing procedure facilitated by a balanced beamsplitter ($50:50$), which effectively erases information about their source or path. Subsequently, the photons are detected using photon number-resolving detectors~\cite{PhysRevLett.91.110405}. The detection of specific click patterns in the detectors confirms the successful generation of an entangled state between the corresponding ions. The information about the parity of the entangled state is obtained
from the photonic detection outcome.

In a recent work by Dhara et al. \cite{PhysRevResearch.5.033149}, the complexities arising from various non-idealities in the entanglement swapping involving photonic qubits were considered, taking into account hardware and channel imperfections. The imperfections include excess noise in the channel and detectors, imperfect mode matching, and carrier-level phase mismatch of the bosonic modes, all of which are practically relevant. In the absence of imperfections, such as optical transmission loss and with the assumption of ideal hardware and number-resolving detectors where $\eta_A = \eta_B = 1$ (they represent the transmissivity of each channel from source $A$ and $B$ to the entanglement swapping station), the linear optical entanglement swap achieves a maximum success probability of 50\%, regardless of the chosen encoding. However, when loss is introduced, the behavior of the entanglement swap changes notably between single rail and dual rail photon qubit encodings. In situations where there is minimal loss in the channel, the dual-rail encoding performs better than the single rail approach. However, as the loss rate rises substantially, the dual-rail encoding suffers a quadratic degradation in performance compared to the single rail encoding.

We consider a specific non-ideal scenario from Ref.~\cite{PhysRevResearch.5.033149} for our analysis in the preparation of remote Bell pairs. The non-idealities include,
\begin{itemize} 

    \item Excess Noise: The system is susceptible to additional noise originating from various sources, including background photons in the channel, noise in detectors, and detector dark clicks. To comprehensively characterize these effects, a parameter  $P_d$, representing the excess photons per mode is defined in Ref.~\cite{PhysRevResearch.5.033149}. The typical range for $P_d$ is within the interval $[0,1)$. When $P_d$ takes a non-zero value, the resulting quantum state becomes mixed, incorporating additional terms whose proportions are approximately of the order of $P_d$ or higher.

    \item Imperfect mode matching: Proper mode matching is vital for effective photonic entanglement swapping, ensuring post-interference analysis conceals path information. Imperfect matching introduces distinguishability among interacting photons, impacting entanglement swapping. Mode mismatch is quantified by visibility parameter $\mathcal{V} \in [0,1]$, where $\mathcal{V}=1$ denotes perfect matching. This parameter represents overall mismatch in the spatio-temporal-polarization mode of the pair of photonic qubits undergoing interference and detection in the optical entanglement swap operation.
\end{itemize}

Among the dual and single rail photonic qubit-based modalities, we consider dual rail qubits as they are better suited for small distance scales relevant to modular architectures for quantum computers. 
We consider the effect of excess noise in the heralded ion-ion entangled state when mediated by the dual rail photonic qubit entanglement swap operation in the symmetric case, i.e., $\eta_A=\eta_B=\sqrt{\eta}$, which are the channel transmissivities from two neighboring sites to a virtual node in the middle where the photonic Bell state measurement is carried out. Assuming perfect detection efficiency of photon detectors and perfect visibility, the complete ion-ion imperfect Bell pair can be represented in terms of the Bell states \(\left\{\left|\Psi^{ \pm}\right\rangle,\left|\Phi^{ \pm}\right\rangle\right\}\) as,
\begin{equation}
\rho_{AB} = \beta_1\left|\Psi^{+}\right\rangle\left\langle\Psi^{+}\right| + \beta_2 \mathbb{I}_4,
\label{rho_AB}
\end{equation}
where \(\mathbb{I}_4\) is the 4-dimensional identity operator, and
\[
\begin{aligned}
    \beta_1 &= \frac{\left(1-P_d\right)^4 \times \eta}{2 \mathbf{N}_d}, \\
    \beta_2 &= \frac{P_d\left(1-P_d\right)^2}{\mathbf{N}_d}\nonumber\\
    &\times \left(\frac{1}{2}\left(1-P_d\right) \sqrt{\eta}(1-\sqrt{\eta})+P_d(1-\sqrt{\eta})^2\right),
\end{aligned}
\]
 \(\mathbf{N}_d\) ensures \(\operatorname{Tr}\left(\rho_{A B}\right)=1\) and equals,
\begin{align}
    N_{d} &= \frac{\left(1-P_d\right)^4 \times \eta}{2} \nonumber\\
    &+ P_d\left(1-P_d\right)^2\left(\frac{1}{2}\left(1-P_d\right) \sqrt{\eta}(1-\sqrt{\eta}) \right. \nonumber\\
    &+ \left. P_d(1-\sqrt{\eta})^2\right).
\end{align}
We can substitute 
\begin{equation}
\mathbb{I}_4 = \left|\Psi^{+}\right\rangle\left\langle\Psi^{+}\right| + \left|\Psi^{-}\right\rangle\left\langle\Psi^{-}\right| + \left|\Phi^{+}\right\rangle\left\langle\Phi^{+}\right| + \left|\Phi^{-}\right\rangle\left\langle\Phi^{-}\right|
\end{equation}
in $\rho_{AB}$,
\begin{equation}
\begin{split}
 \rho_{A B} = (\beta_1+\beta_2)\left|\Psi^{+}\right\rangle\left\langle\Psi^{+}\right| + \beta_2 \left|\Psi^{-}\right\rangle\left\langle\Psi^{-}\right| + \\
 \beta_2 \left|\Phi^{-}\right\rangle\left\langle\Phi^{-}\right|+
 \beta_2 \left|\Phi^{+}\right\rangle\left\langle\Phi^{+}\right|.
 \end{split}
\end{equation}

% \begin{figure*}[t]%[t]%[t] % Use figure* instead of figure
%   \centering
% \includegraphics[width=\textwidth]{contour}
%   \caption{ The contour plots show the value of depolarizing noise for different values of excess photons per mode $P_{d}$ from $10^{-5}$ to $10^{-1}$ and half-channel transmissivity $\sqrt{\eta}$  from $0$ to $1$ for $T = 100 \mu s$ and $\tau_{D} = 1\text{s}$ \label{contour}.
% The left contour plot shows the value of depolarizing noise parameter in the entire considered parametric regimes of $P_{d}$ and $\sqrt{\eta}$. The right contour plot shows values of the parametric regimes within the noise tolerance threshold.}
%   \label{fig:wide-figure}
% \end{figure*}

% Our goal is to see if we can extract a depolarizing noise parameter from the above equation and if successful, find the relevant depolarizing noise parameter. 
Our goal is to extract the depolarizing noise parameter from the equation, which we obtain, as demonstrated below. We consider an ideal Bell state, acted on by a single-qubit depolarizing noise channel ($\xi$)  independently on both the qubits, 
\begin{equation}
    \xi_{i}^{\delta} : \rho_{i}\rightarrow \delta \rho_{i} + (1-\delta) \frac{\textbf{I}_{2}}{2}
\end{equation}
where $\rho_{i}$ is the density matrix of the single qubit acted upon by the $i^{th}$ single qubit depolarizing channel ($\xi_{i}^{\delta}$) and $\delta$ is the depolarizing noise parameter. The action of this channel on $\left( \rho = |\psi^{+}\rangle \langle\psi^{+}| \right)$ is given by,

% \begin{equation}
%     \xi_{1}^{\delta} \otimes\xi_{2}^{\delta} (\rho = |\psi^{+}\rangle\langle\psi^{+}|) = \delta^2 \rho +\delta(1-\delta) Tr_{2} (\rho)\otimes \frac{\textbf{I}_2}{2} 
%     \\ + \delta(1-\delta) \frac{\textbf{I}_2}{2}\otimes Tr_{2} (\rho) + (1-\delta)^{2}\frac{\textbf{I}_4}{2}
 
% \end{equation}

\begin{equation}
\begin{aligned}
    \xi_{1}^{\delta} \otimes \xi_{2}^{\delta} \left( \rho  \right) = & \, \delta^2 \rho 
    + \delta(1-\delta) \, \text{Tr}_{2}(\rho) \otimes \frac{\textbf{I}_2}{2} \\
    & + \delta(1-\delta) \, \frac{\textbf{I}_2}{2} \otimes \text{Tr}_{2}(\rho) \\
    & + (1-\delta)^{2} \frac{\textbf{I}_4}{2}
\end{aligned}
\end{equation}
which simplifies to,
\begin{equation}
    = \delta^{2} \rho + (\frac{1-\delta^{2}}{4}) \textbf{I}_{4}
\end{equation}
% We know that sum of all four Bell pairs is $I_{4}$ a $4\times4$, Identity matrix. 
Comparing the above equation with Eq.~(\ref{rho_AB}), we find,

\begin{equation}
\begin{aligned}
    \beta_{1} = \delta^{2}, \\
    \beta_{2} = \frac{(1-\delta^{2})}{4}.
    \end{aligned}
\end{equation}
We see that $\beta_{1} + 4\beta_{2} = 1$, satisfying the normalization condition of the quantum state. 

Equivalently, we can define the single qubit depolarizing channel in terms of the error probability $\frac{p}{3}$, where $\frac{p}{3}$ is the probability of occurrence of each of the Pauli errors ($X$,$Y$ and $Z$) on the qubit as,

\begin{equation}
    \xi^{p}(\rho) = (1-p)\rho + \frac{p}{3}(X\rho X + Y\rho Y + Z\rho Z) \label{dp}
\end{equation}
where $p = \dfrac{3}{4}(1-\delta).$ We can demonstrate the equivalence of these two definitions of the depolarizing channel by expressing the channel in the operator-sum representation and using the completeness relation for \( \rho \in \mathcal{D}_{2} \) (the set of \( 2 \times 2 \) density matrices)~\cite{nielsen2001quantum}. This leads to the following property,
\begin{equation}
    \dfrac{1}{4}[1\rho 1 + X\rho X + Y\rho Y + Z\rho Z] = \dfrac{\textbf{I}_{2}}{2}.
\end{equation}
If we substitute $\delta = 1- \dfrac{4p}{3}$ in $\xi_{\rho}^{\delta}$ and use the above property, we get,
\begin{equation}
\xi^{\delta}_{i} = (1-p)\rho + \dfrac{p}{3}(X\rho X + Y\rho Y + Z\rho Z)
\label{eqn_same}
\end{equation}
which is same as Eq.~(\ref{dp}).

% Assuming an initial state $\left|\psi^{+}\right\rangle_{A B}$ which is the ideal intended preparation, when a depolarizing channel acts on qubit $B$, the entangled state evolves as~\cite{preskill2015lecture}
% \begin{align}
% \left|\psi^{+}\right\rangle\left\langle\psi^{+}|\mapsto(1-p)| \psi^{+}\right\rangle\left\langle\psi^{+}\right|+\nonumber\\
% \frac{p}{3}\left(\left|\phi^{+}\right\rangle\left\langle\phi^{+}|+| \psi^{-}\right\rangle\left\langle\psi^{-}|+| \phi^{-}\right\rangle\left\langle\phi^{-}\right|\right)
% \end{align}
% where probability $p$ is associated with the occurrence of an error, and the errors be any of three Pauli errors ($X$, $Y$, and $Z$), each with an equal likelihood. From equating the above two states,
% \begin{equation}
% \begin{split}
%     \beta_{1} = 1-\frac{4p}{3} \\
%     \beta_{2} = \frac{p}{3} 
% \end{split}
% \end{equation}
% Alternatively, the state after the action of depolarizing noise can be written as,
% \begin{equation}
%     (1-\frac{4p}{3})|\psi^{+}\rangle\langle\psi^{+}| +\frac{4p}{3}\frac{I_{4}}{4}
% \end{equation}
% The action of quantum depolarising channel on initial state $\rho = |\psi^{+}\rangle\langle\psi^{+}|$ can also be described as,
% \begin{equation}
%     \rho \mapsto (1-\epsilon)\rho + \dfrac{\epsilon}{4}I_{4}
% \end{equation}
% From equating the above two equations, we find the depolarizing parameter is given by
% $\epsilon = \dfrac{4p}{3} = 4\beta_{2}(P_{d},\eta)$.

\subsection{Analysis of Noise Tolerance}

In order to analyze the noise tolerance of the RHG lattice within our architecture, we briefly summarize the steps involved in the generation (see subsection \ref{rhg_generation}). There are primarily three fundamental steps: (i) Generating ion-ion Bell pairs between distinct ELUs through photonic mediation, (ii) Implementing CNOT gates within each ELU, and (iii) Performing local measurements on qubits in each ELUs.

% We consider the following error model. (1) All gate operations, including the preparation and measurement of individual qubits, gates within an ELU, and the creation of Bell statepairs between distinct ELUs through the photonic link, (2) implementing CNOT gates within each ELU, and fferent ELUs, can be completed within a clock cycle of duration $T$. An erroneous one-qubit (two-qubit) gate is represented by the perfect gate followed by a partially depolarizing one-qubit (two-qubit) channel. In the one-qubit channel, errors $X, Y$, and $Z$ each occur with a probability of $\epsilon / 3$. In the two-qubit channel, each of the 15 possible errors $X_1, X_2, X_1 X_2, \ldots, Z_1 Z_2$ occurs with a probability of $\epsilon / 15$. All gates have the same error rate $\epsilon$. (32) conducting local measurements on  qubits in each ELU.Additionally, the impact of decoherence per time step $T$ is described by local probabilistic Pauli errors $X, Y, Z$, each occurring with a probability of $T / 3 \tau_D$ where $\tau_{D}$ is the decoherence time of the qubits.

The assumptions and error model associated with the different stages are,
\begin{itemize}
  \item \textit{Gate Operations}: Each of the operations, including the establishment of Bell state pairs between distinct ELUs via photonic mediation, gate executions within an ELU, and individual qubit measurements, can be completed within a single time step of duration \( T \).

     % All gate operations, encompassing the preparation and measurement of individual qubits, the execution of gates within an ELU, and the establishment of Bell state pairs between distinct ELUs via the photonic link, can be accomplished within a single clock cycle of duration $T$.
    
\item {\textit{Bell State Preparation}:} 
We account for the excess noise in the heralded ion-ion entangled state arising from the remote entanglement generation involving realistic single-photon dual-rail qubits and their detection as part of the optical entanglement swap operation. The ensuing imperfect Bell pair is modeled as a perfect Bell pair subjected to a single-qubit depolarizing channel on both qubits.

% We consider the excess noise present in the heralded ion-ion entangled state when it is mediated by the dual-rail photonic qubit entanglement swap operation, resulting in an imperfect Bell pair. This imperfect Bell pair can be considered as a perfect Bell pair subjected to a single qubit depolarizing channel impacting each of the qubits.
  
  % \item {Erroneous Gates and Error Probabilities :} An erroneous one-qubit (two-qubit) gate is represented by the perfect gate followed by a partially depolarizing one-qubit (two-qubit) channel. In the one-qubit channel, errors $X, Y$, and $Z$ each occur with a probability of $\epsilon / 3$. In the two-qubit channel, each of the 15 possible errors $X_1, X_2, X_1 X_2, \ldots, Z_1 Z_2$ occurs with a probability of $\epsilon / 15$, where all gates have the same error rate $\epsilon$.
  
  \item 

 \textit{Erroneous Gates and Error Probabilities}: We represent an erroneous one-qubit (or two-qubit) gate as a perfect gate followed by a partially depolarizing one-qubit (or two-qubit) channel. In the one-qubit channel, errors \(X\), \(Y\), and \(Z\) occur with a probability of \(\dfrac{\epsilon}{3}\) each. In the two-qubit channel, each of the 15 possible errors—such as \(X_1\), \(X_2\), \(X_1 X_2\), ... and  \(Z_1 Z_2\)—occurs with a probability of \(\dfrac{\epsilon}{15}\), with all gates having the same error \(\epsilon\).

  \item \textit{Decoherence}: The effect of decoherence at each time step \( T \) is represented by local probabilistic Pauli errors \( X \), \( Y \), and \( Z \), each occurring with a probability of \( \dfrac{T}{3 \tau_D} \), where \( \tau_D \) denotes the decoherence time of the qubits.

\end{itemize}

We derive the noise tolerance threshold for the described error model, assuming an absence of bond failure errors (see Appendix~\ref{appendix}). The threshold is defined in terms of an inequality criterion established for the error threshold of measurement-based quantum computation using cluster states. It is based on numerical results obtained for various error models~\cite{KITAEV20032,RAUSSENDORF20062242} and considers the error contributions arising from a limited number of error sources. The derived inequality is given by,
\begin{equation}
    \epsilon + \frac{35}{16} \frac{T}{\tau_D} + \frac{35}{64} (1 - \sqrt{\beta_{1}(\eta,P_{d})}) < 3.9\times 10^{-3} .
    \label{ft_equation}
\end{equation}
% where $T=\tau_2$, i.e., the time for creation of the two layers of the unit cell of the RHG lattice.  %\kaushik{In MUSIQC, as I understand, $T$ is in fact the time to create the full RHG lattice unit cell. However, in the bilayer implementation, I notice that we not only create half unit cell first, but also measure first layer before completing the unit cell creation. So, question is do you think we can reconcile the use of this inequality from the MUSIQC paper as is, or should we modify it accordingly? If so, how?}

% We employ the above inequality of to examine i) the parametric regime of detector efficieny ($\eta$)  and Excess Noise $P_{d}$ representing excess photons per mode that lie within the fault tolerance regime of our architecture for a small depolarizing noise of $\epsilon = 10^{-4}$. It is the surface above the red line which falls under fault tolerance regime. 

\begin{figure*}[t]%[t]%[t] % Use figure* instead of figure
  \centering
\includegraphics[width=13cm]{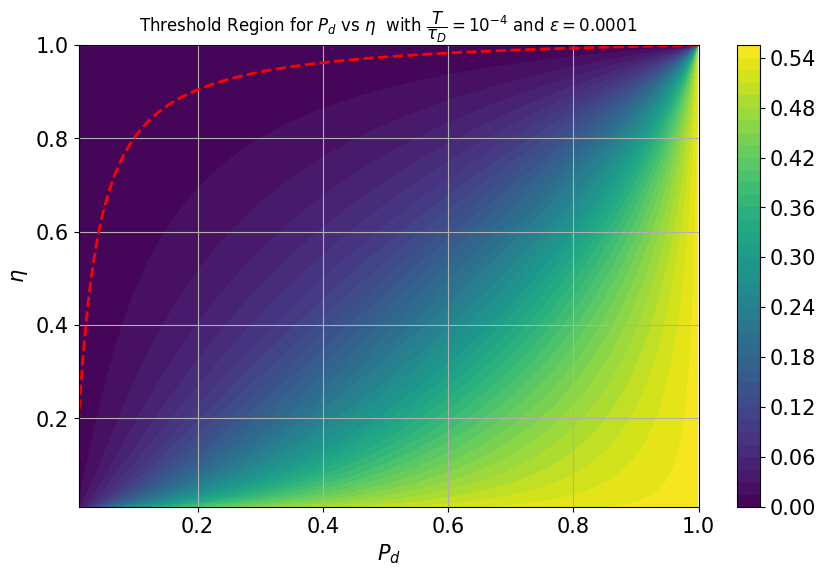} 
  \caption{ The parametric regime of channel transmissivity \(\eta\) and excess noise \(P_d\), which corresponds to the excess photons per mode that fall within the fault tolerance (FT) regime of our architecture lies above the red line for a small gate error of \(\epsilon = 10^{-4}\). The color gradient in the contour plot reflects the values obtained by evaluating Eq.~\ref{ft_equation} for various parameters.
}
  \label{plot1_ft}
\end{figure*}

% \section{Fault Tolerant Calculations for our model}

\begin{figure*}[t]%[t]%[t] % Use figure* instead of figure
  \centering
\includegraphics[width=13cm]{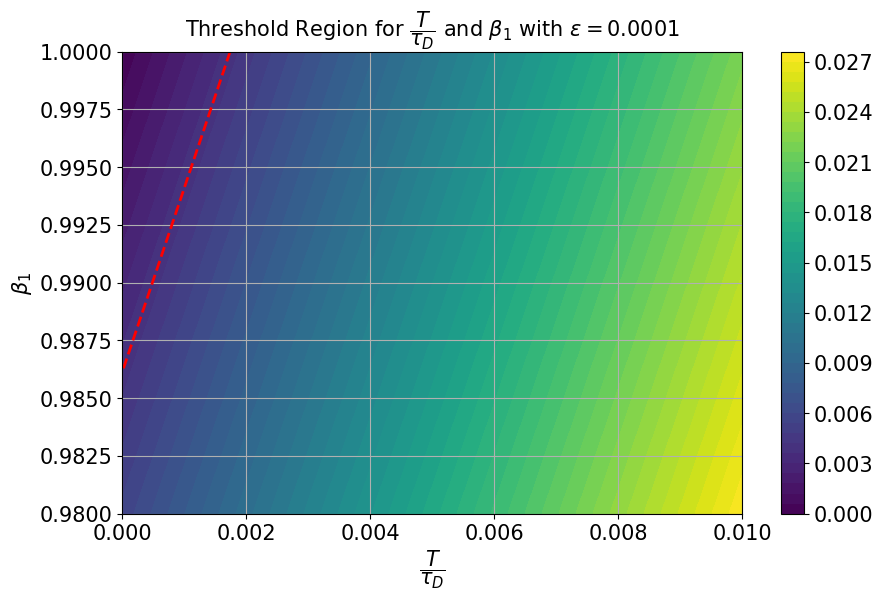} 
  \caption{ The parametric regime of depolarizing noise parameter \(\beta_1\) and the ratio of duration of the
timestep (See Eq.\ref{timestep}) to qubit decoherence time \(\frac{T}{\tau_D}\), for a small gate error of \(\epsilon = 10^{-4}\)  above the red line corresponds to a small region in the plane that satisfies the fault-tolerant inequality. The color gradient in the contour plot reflects the values obtained by evaluating Eq.~\ref{ft_equation} for various parameters.
}
  \label{fig_ft2}
\end{figure*}

\begin{figure*}[t]%[t]%[t] % Use figure* instead of figure
  \centering
\includegraphics[width=13cm]{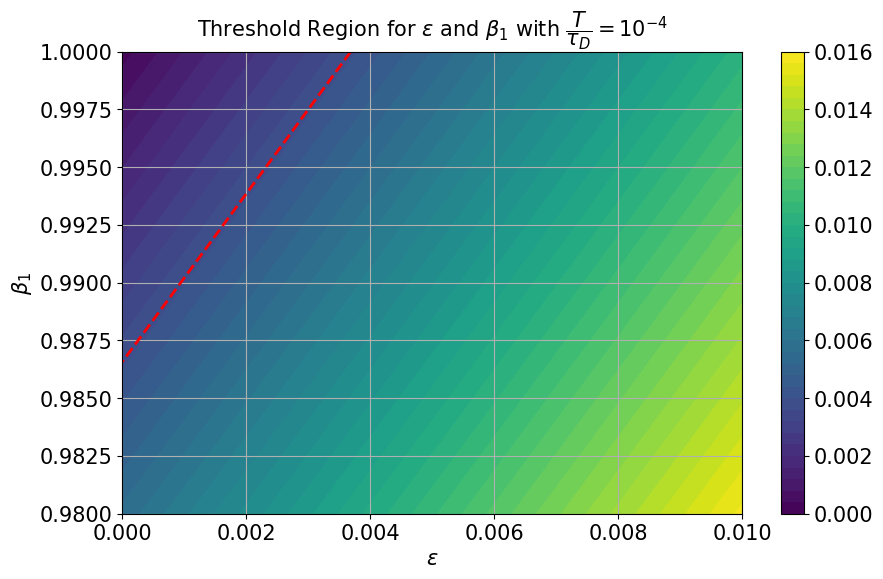} 
  \caption{ The parametric regime of depolarizing noise parameter \(\beta_1\) and gate error \(\epsilon\) for a fixed \(\frac{T}{\tau_D} = 10^{-4}\) above the red line in the \((\epsilon, \beta_1)\) plane indicates the region that satisfies the fault-tolerance inequality. The color gradient in the contour plot reflects the values obtained by evaluating Eq.~\ref{ft_equation} for various parameters.
 \label{contour_ft3}
}
  \label{fig:wide-figure2}
\end{figure*}

We utilize the above inequality to investigate the following:

\begin{itemize}
    \item[$(i)$] The parametric regime of channel transmissivity \(\eta\) and excess noise \(P_d\), which represents the excess photons per mode that fall within the fault tolerance (FT) regime of our architecture for a small gate error of \(\epsilon = 10^{-4}\). The area above the red line in Fig.~\ref{plot1_ft} corresponds to the fault tolerance regime.

    \item[$(ii)$] The parametric regime of \(\beta_1\) (a parameter describing depolarizing noise) and the ratio of a single time step duration to decoherence time of the qubits \(T/\tau_{D}\), for a small gate error of \(\epsilon = 10^{-4}\). From Fig.~\ref{fig_ft2}, we observe a small region in the \(\left(T/\tau_{D}, \beta_1\right)\) plane that lies above the red line, indicating that it satisfies the FT-inequality.

    \item[$(iii)$] The parametric regime of \(\beta_1\) and gate error \(\epsilon\) for \(T/\tau_{D} = 10^{-4}\), which meets the FT-inequality. In Fig.~\ref{contour_ft3}, a small region in the \((\epsilon, \beta_1)\) plane lies above the red line, indicating that it resides within the FT-regime.
\end{itemize}

Through this analysis, we identify specific parameter values for various noise factors—such as channel transmissivity (\(\eta\)), excess photons per mode (\(P_d\)), depolarizing noise parameter (\(\beta_1\)), and gate error (\(\epsilon\))—that allow the system to operate within the fault tolerance (FT) regime. The plots  demonstrate that the quantum system can function in the FT regime only within specific ranges of these parameters.

The excess photons per mode
represented by the parameter \(P_d\) serves as an indicator of the combined effects of several noise sources, including background photons in the communication channel, electronic Johnson-Nyquist noise in the detectors, and detector dark counts. These noise sources are intrinsically linked to the channel transmissivity value, \(\eta\). For the case of near-perfect channel transmissivity, where \(\eta\) is close to 1, almost the entire range of excess photons per mode, \(P_d\), lies within the FT threshold.  However, as \(\eta\) decreases, corresponding to increased channel losses, we observe a significant reduction in the permissible range of \(P_d\). This reduction indicates that the system becomes more sensitive to excess noise as the channel transmissivity degrades, thereby shrinking the region where the system can operate fault-tolerantly.

Furthermore, we observe that only a very narrow range of parameter describing the depolarizing noise, \(\beta_1\), falls within the FT regime. This occurs when the ratio of the duration of time step while preparing the RHG lattice to the qubit decoherence time, \(T/\tau_{D}\), is kept very small. The situation is further complicated by the gate error, \(\epsilon\), which must also remain extremely small for the system to stay within the FT regime. Therefore, for a given small value of \(T/\tau_{D}\), the allowable range of depolarizing noise \(\beta_1\) is extremely limited, further constraining the system's operational window in the FT regime.

% when the average entanglement
% creation time $\tau_{E}$ is considerably smaller than the decoherence
% time scale $\tau_{D}$.

% Putting the value in the noise tolerance threshold inequality, we have,.....
% \begin{equation}
%     4\beta_{2}(P_{d},\eta) + \frac{55}{32} \frac{T}{\tau_D} < 2.9 \times 10^{-3}.
% \end{equation}

% In order to satisfy this inequality, $\eta$ and $P_{d}$ can take only a specific set of values. Figure~\ref{contour} shows the parametric regimes satisfying the inequality using  contour plots for representative values of $T$ and $\tau_D$.

% \begin{figure}%[H]
%        \centering
% \includegraphics[width=8cm]{eta_plots.jpg}
%         \caption{Depolarizing error parameter $\epsilon$ vs excess photons per mode $P_d$ with threshold line and shaded region for different values of $\eta$, $T=100 \mu s$, $\tau_{D}=1s$.}
%          \label{error1}
%     \end{figure}

% \begin{figure}%[H]
%        \centering
% \includegraphics[width=8cm]{Pd_plots.jpg}
%         \caption{Depolarizing error parameter 
%  $\epsilon$ vs channel transmissivity $\sqrt{\eta}$ with threshold line for different values of excess photons per mode $P_{d}$, $T=100 \mu s$, $\tau_{D}=1s$.}
%          \label{error2}
%     \end{figure}

% Figure~\ref{error1} shows depolarizing error parameter $\epsilon$ as a function of excess photons per mode $P_d$. 

\section{Conclusion}\label{conc}
We investigated an architecture for fault-tolerant measurement-based quantum computation (FT-MBQC) using optically networked dual-species trapped-ion (DSTI) modules. This design employs photonic interactions between modules and Coulombic interactions within modules, with the number of modules matching the number of sites in two layers of the lattice for the topologically protected resource cluster state. To improve the success rate of Bell pair generation between distant modules, we utilized spatial and temporal multiplexing. We calculated the time required for one code cycle in the RHG cluster state and the resources needed to achieve any target code cycle duration. For large inter-site distances, we introduced quantum repeaters and compared the resource requirements, with and without repeaters, as a function of distance. We derived an inequality relating the system's parameters to ensure fault tolerance. Additionally, we analyzed the architecture's robustness to noise from imperfect Bell pairs generated via dual-rail photonic qubits, accounting for both hardware and channel imperfections. By comparing this noise to the RHG lattice’s fault tolerance threshold, we identified the optimal system parameters to maintain fault tolerance. Our work addresses both bond failure and noise tolerance in the RHG lattice for optically networked trapped-ion systems and contributes to the growing literature on practical distributed architectures for realizing FT-MBQC. 

\section{Acknowledgements}
This work was supported by NSF award 2134891 QuaNeCQT project (Quantum Networks to Connect 
Quantum Technology) and NSF award 2204985. NKC and KPS thank Prajit Dhara and Eneet Kaur for helpful discussions. KPS thanks Norbert Linke, Nithin Raveendran, Narayanan Rengaswamy, Tripti Sinha and Edo Waks for helpful discussions.

\section{Appendix: ERROR PROBABILITY CALCULATION}\label{appendix}

In this appendix, we provide a detailed calculation of the total error probability associated with the stabilizer measurement process, based on the error model considered in our paper. This model accounts for various sources of noise, including imperfections in remote Bell state preparation between trapped ion modules, and gate errors in both single-qubit and two-qubit operations, characterized by a gate error probability \(\epsilon\). Additionally, the effects of qubit decoherence are considered as a single qubit Pauli error whose total probability is modelled as the ratio between the time step duration and the qubit decoherence time, \(T/\tau_{D}\) (assumed equally likely to be an $X$, $Y$ or $Z$ error).

The stabilizer operator expectation value, denoted as \(\left\langle K_{\partial q} \right\rangle\), reflects the cumulative impact of these errors. Mathematically, it is expressed as:
\begin{equation}
\left\langle K_{\partial q} \right\rangle = \prod_{E \in \text{error sources}} \left( 1 - 2p_E \right),
\end{equation}
where \(p_E\) represents the total probability of Pauli errors originating from the error source \(E\). These errors, when accrued and propagated through the cluster state generation process, can interfere with the stabilizer measurement by anticommuting with the stabilizer operator \(K_{\partial q}\). The product form of the above expression results from the assumption that different error sources are statistically independent, meaning their contributions to the overall error probability can be treated as independent events.

% The method used for generating the cluster state plays a critical role in the propagation of errors. Because the process is confined to a specific temporal depth and involves only local and nearest-neighbor quantum gates, the spread of errors is limited to a finite distance. This means that only a small number of error sources affect the total error probability at any point in the procedure.

Once we have computed the expectation value of \(\left\langle K_{\partial q} \right\rangle\) for the relevant error parameters, we compare the result against a known threshold value. Previous numerical studies on measurement-based quantum computation (MBQC) with cluster states have established an empirical error threshold criterion for various error models~\cite{RAUSSENDORF20062242,WANG200331}. To ensure the system operates within the fault-tolerant regime, the stabilizer expectation value must satisfy the following condition:
\[
\left\langle K_{\partial q} \right\rangle (\{\text{error parameters}\}) > 0.70.
\]
This threshold can be used to establish an inequality involving the system's error parameters and a constant, which must be satisfied to ensure the system remains within the fault-tolerant regime, despite the presence of noise and gate errors.

The error sources contributing to \(\langle K_{\delta q} \rangle\) can be grouped into three distinct categories:

\begin{itemize}
    \item Type-I: Errors occurring during the creation of the first Bell pair on each face of the elementary logical units (ELUs). These are the qubits that live on to eventually be part of the RHG lattice.
    \item Type-II: Errors from teleported CNOT gates as part of the RHG lattice generation that are accomplished by consuming the remaining ancillary Bell pairs.
    \item Type-III: Errors introduced during the final measurement of the qubits in the cluster state.
\end{itemize}

% This threshold leads to an inequality involving error parameters and a constant required for the system to maintain fault tolerance, in the presence of noise and gate errors.

% We have three categories of error sources that 
% contribute to $\langle K_{\delta q}\rangle$.
% \begin{itemize}
%     \item Type-I : Creating the first Bell pair on each face ELUs.
%     \item Type-II : Teleported CNOT links that consume the remaining Bell pairs.
%      \item Type-III : Final measurement of cluster state qubits.
% \end{itemize}
     \subsection{Type-I: Initial Bell pair creation}
     The error propagation in this step can be detailed as follows:
\begin{itemize}
    \item First of all, in generating the RHG cell, we only consider the preparation of Bell pairs on three of its faces. The remaining  faces will be accounted for by neighboring cells. %Therefore, we focus on Bell pairs of three faces of a unit cell.
    \item In the preparation of a remote Bell pair, we model the error as a single qubit depolarizing noise of strength $\delta$ acting independently on both the qubits of the Bell pair.
\item We consider single qubit memory errors acting independently on each qubit as depolarizing noise of strength $T/\tau_{D}$, where $T$ is the duration of a time step (See Eq.~\ref{timestep}) and $\tau_{D}$ is the qubit coherence time.
\end{itemize}

The stabilizer $K_{\delta q}$ is supported only by the face qubits, and since it is a $X$-type stabilizer, it is unaffected by \( X \) errors. There are $4$ possible single-qubit depolarizing errors \{$Y_{e},Z_{e},Y_{f},Z_{f}$\} (where subscripts $e$ and $f$ denote qubit location, namely, an edge node or a face node, respectively) as only $Z$ errors can affect the stabilizer (which includes $Y$ errors, since  $Y=iXZ\simeq Z$, and $X$ errors commute with the cluster state unit cell stabilizer). Each of these errors occur with probability $p/3$. Here $p = 3(1-\delta)/4$ (See Eq.~(\ref{eqn_same})) for the noise from the imperfect remote optical Bell state generation, and $T/\tau_{D}$ for the noise due to decoherence in a time step. Therefore, $p_{ZI}$, representing a $Z$ error on the face qubit and no error on the edge qubit of the Bell pair, due to the sum of single qubit depolarizing noise from imperfect remote preparation and memory decoherence is,
\begin{equation}
    p_{ZI} = \dfrac{4}{3}\left(\dfrac{3}{4}(1-\delta) + \frac{T}{\tau_{D}}\right)
  \end{equation}  
which simplifies to,
\begin{equation}
     p_{ZI} = (1-\delta) + \dfrac{4}{3}\dfrac{T}{\tau_{D}}.
     \label{bell_memory}
\end{equation}
 Since the number of initial Bell pairs is three, the total type-I error that contributes to $\langle K_{\delta_{q}}\rangle $ is given by, 

\begin{equation}
    \quad=\left[1-\frac{8}{3}\left(\frac{3}{4}(1-\delta)+\frac{T}{\tau_D}\right)\right]^3
    \end{equation}
 \begin{equation}
\quad \approx 1-6(1-\delta)-8 \frac{T}{\tau_D} \quad .
\end{equation}
\hspace{60pt} (Upto linear approximation)

% (1) 2-Qubit Pauli Eros in Preparation
% - 15 possibilities
% - Each $w /$ probability $E / 15$
% - 3 equivalena classes.

% $$
% \begin{array}{rl}
% * & I:\left(z_e z_f, X_e x_f, y_e y_f\right), \mid \pm l=3 \\
% * & Z_f:\left(\text { Not }: Z_f=z_e \text { for Bell state }\right) \\
% & \left(Z_e, Z_f, X_e y_f, y_e x_f, y_e, y_f, X_e z_f, x_f z_e\right) \\
% & \left|Z_f\right|=8 \\
% * & x:\left(x_e, x_f, y_e z_f Z_e y_f\right),|x|=4
% \end{array}
% $$

\subsection{Type II: Contributions from teleported CNOT links}

\begin{figure}%[H]
       \centering
 \includegraphics[width= 9.5 cm]{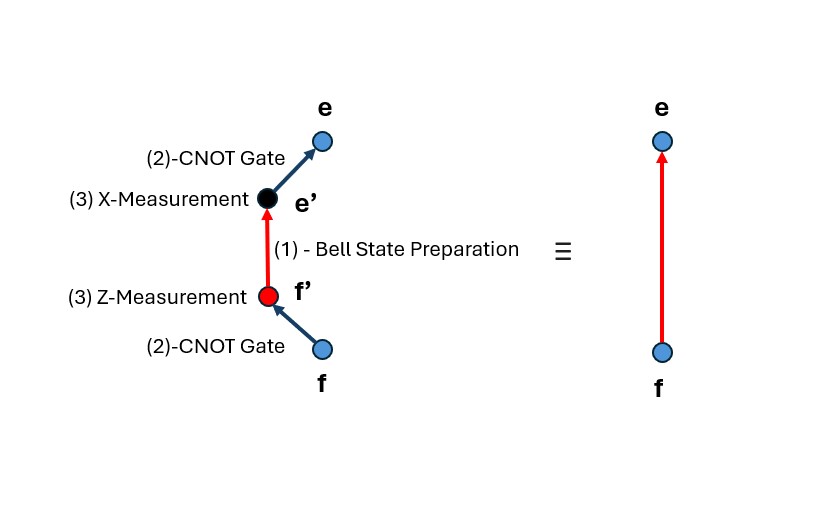}
        \caption{A teleported CNOT between $e$ and $f$ qubits using a Bell pair involving $e^{'}$ and $f^{'}$ qubits and two CNOT gates between $f-f^{'}$ \text{and} $e^{'}-e$ respectively.}
        \label{fig:tele}
    \end{figure}

We now describe the errors accrued and propagated via the teleported CNOT gates that are effectively accomplished between qubits (from control at a face node to target at an edge node) in the RHG lattice cluster state along the procedure of Sec.~\ref{rhg_generation}. Consider the sequence of operations described in Fig.~\ref{fig:tele}. Error propagation along this process can be outlined in the following steps:
\begin{enumerate}%[label=(\alph*)]
    \item Creation of the additional ancilla Bell pairs, which are later consumed in generating the teleported CNOT links, along with the first round of memory errors affecting all qubits.
    \item Execution of the CNOT gates, followed by the second round of memory errors on all qubits.
    \item The final step involves performing \(X\) and \(Z\) measurements on the Bell pairs.
\end{enumerate}

When examining a CNOT link, we focus solely on $Z$ errors (which includes $Y$ errors too as $Y=iXZ\simeq Z$) that occur on both the control qubit (the face) and the target qubit (the edge). This is because $Z$ errors on the face qubit do not commute with the stabilizer $K_{\delta q}$, which is supported on face qubits, and of $X$-type (and therefore unaffected by $X$ errors). Additionally, $Z$ errors on the edge qubit can affect a neighboring face qubit via the teleported CNOT gate, since error propagation of Pauli errors ($X$ and $Z$) over a CNOT gate are given by
\begin{align}
& X_c \rightarrow X_c X_t \\
& X_t \rightarrow X_t \\
& Z_c \rightarrow Z_c \\
& Z_t \rightarrow Z_c Z_t.
\end{align}
where $c$ refers to control qubit and $t$ refers to target qubit. 
We note that errors can propagate only once between a face qubit and an edge qubit, and not beyond that. For instance, an $X$ or $Y$ error on a face qubit can be transferred (as the face acts as the control for the CNOTs). The overall error associated with each teleported CNOT link between two neighboring ELUs can be represented using three probabilities: $p_{Z I}$ for a $Z$ error on the face qubit alone, $p_{I Z}$ for a $Z$ error on the edge qubit alone, and $p_{Z Z}$ for a $Z$ error on both qubits.

The errors involved in $p_{IZ}$ come from a CNOT gate $(2)$ and memory error in $e$ qubit (see Fig.~\ref{fig:tele}). 
% From $15$ two-qubit Pauli errors, only errors from the set \{$IZ, IY, XZ, YZ$\} where $P_{e'}P_{e}$ refers to Pauli errors on $e'$ and $e$ respectively  leads to the error of $4\dfrac{\epsilon}{15}$. The other error comes from $Y$ and $Z$ memory errors contributing $\dfrac{2T}{3\tau_{D}}$ in $p_{ZI}$. Thus,
Out of the 15 possible two-qubit Pauli errors, only errors from the set \{$IZ, IY, XZ, YZ$\}, where \( P_{e'}P_{e} \) refers to Pauli errors on \( e' \) and \( e \), respectively, contribute to an error of \(\epsilon/15 \) each. Additionally, \( Y \) and \( Z \) memory errors contribute an error term of \( 2T/(3\tau_{D}) \) in \( p_{ZI} \).
\begin{equation}
    p_{IZ} = 4\dfrac{\epsilon}{15} + \dfrac{2T}{3 \tau_{D}}.
\end{equation}

The errors involved in $p_{ZI}$ come from a) Bell pair creation and the memory errors, b) Error propagation by the CNOT gate and the erroneous two-qubit CNOT gate, and c) Performing $X$ and $Z$ measurements on the qubits.
The errors involved in step (1) $p_{ZI}^{a}$ is given by,
\begin{equation}
    p_{ZI}^{a} = (1-\delta) + \dfrac{4T}{3\tau_{D}} \quad\text{  (See Eq.~(\ref{bell_memory}))}.
\end{equation}
The errors involved in step (2) $p_{ZI}^{b}$ is given by,
\begin{equation}
    p_{ZI}^{b} = 12\left(\dfrac{ \epsilon}{15}\right) + 6\left(\dfrac{T}{3\tau_{D}}\right).
\end{equation}
The first term comes from two-qubit Pauli errors \{$IZ,ZI,IY,YI,XZ,ZX,XY,YX$\} where $P_{f'}P_{f}$ are two-qubit Pauli errors on $f'$ and $f$, respectively, and \{$ZI,YI,ZX,YX$\} where $P_{e'}P_{e}$ are two-qubit Pauli errors on $e'$ and $e$ respectively. The second term comes from memory errors in qubits involving CNOT gates. 

There are $15$ possible Pauli errors, each occurring with a probability of \(\epsilon/15\). These errors can be grouped into three distinct equivalence classes:

\[
\begin{array}{ll}
\textbf{Class $I$}: & \left( Z_e Z_f, X_e X_f, Y_e Y_f \right), \quad |I| = 3 \\
\\
\textbf{Class $Z$: }  & \text{ (We note that } Z_e \equiv Z_f \text{ for Bell states)} \\
& \left( Z_e, Z_f, X_e Y_f, Y_e X_f, Y_e, Y_f, X_e Z_f, X_f Z_e \right),\\  |Z_f| = 8 \\
\\
\textbf{Class $X$}: & \left( X_e, X_f, Y_e Z_f, Z_e Y_f \right), \quad |X| = 4
\end{array}
\]
Each class can be described as:
\begin{itemize}
    \item Class $I$: Includes errors that act trivially on the Bell state.
    \item Class  \(Z\): Consists of errors where \(Z_e \equiv Z_f\) for Bell states, with 8 possible error configurations.
    \item Class $X$: Contains errors involving \(X_e\), \(X_f\), or a combination of \(Y_e Z_f\) and \(Z_e Y_f\).
\end{itemize}
% The 15 possible Pauli errors can be classified into equivalence classes as follows: \(I\), \(Z_f\), where \(Z_e Z_f \equiv I\) and \(Z_e \equiv Z_f\) for Bell states.
% \begin{aligned}
% & e_e^{\prime}\left(\begin{array}{cccc}
% z & y & z & y \\
% 1 & 1 & x & x
% \end{array}\right)=\left[4 \frac{E}{15}\right]
% \end{aligned}
The errors involved in step (c) comes from the measurement-$(3)$  shown in Fig.~\ref{fig:tele},  $p_{ZI}^{c}$ is given by,

\begin{equation}
    p_{ZI}^{c} = 2\left(\dfrac{ \epsilon}{3}\right) .
\end{equation}
Combining the above three steps, we get,
\begin{equation}
    p_{ZI} = 22\left(\dfrac{ \epsilon}{15}\right) + 10\left(\dfrac{T}{3 \tau_{D}}\right) + \left(1-\delta\right).
\end{equation}

We notice that \(p_{ZZ}^{f-e} = p_{IZ}^{f-e}\) because qubits \(f\) and \(e\) are connected by a teleported CNOT link. As a result, a \(Z\) error on the target qubit \(e\) propagates to the control qubit \(f\), leading to a \(ZZ\) error on the \(f-e\) link. Out of $p_{ZI}$, $p_{IZ}$ and $p_{ZZ}$, only the first two terms contribute in reducing $K_{\delta q}$. Thus,  ($1 - 2(p_{ZI}+p_{IZ})$) or  ($1 - 2(p_{IZ}+p_{ZI})$) simplifies to, 
\begin{align}
   1 - &2\bigg(22\left(\dfrac{ \epsilon}{15}\right) + 10\left(\dfrac{T}{3\tau_{D}}\right)\nonumber\\
   &+ (1-\delta) + 4\left(\dfrac{\epsilon}{15}\right) + 2\left(\dfrac{T}{3\tau_{D}}\right)\bigg).
\end{align}
For every teleported CNOT link, these errors reduce $K_{\delta q}$ by a factor of,
\begin{equation}
    1 - \dfrac{52\epsilon}{15} - \dfrac{8T}{\tau_{D}} - 2(1-\delta).
\end{equation}
The error contribution in $K_{\delta q}$ comes from three links within the cell  and there are six faces. This leads to,

\begin{equation}
    = 1 - 2 \left( \frac{26\epsilon}{15} + \frac{4T}{\tau_{D}} + (1 - \delta) \right)^{18}
\end{equation}

\begin{equation}
   \approx 1 - 36 \left( \frac{26\epsilon}{15} + \frac{4T}{\tau_{D}} + (1 - \delta) \right).
   \label{type_1}
\end{equation}
The error contribution in $K_{\delta q}$ from neighboring cells is three for every cell $\delta q$  and there are six faces. But out of $18$, only $6$ can propagate to an odd number of faces. So, the total error probability ($p_{IZ} + p_{ZI}$) or ($p_{ZZ} + p_{ZI}$) is,
\begin{equation}
   = 8\left(\dfrac{ \epsilon}{15}\right) + 4\left(\dfrac{T}{3\tau_{D}}\right).
\end{equation}
For six links affecting $K_{\delta q}$, the  expression becomes,
\begin{equation}
    = \left( 1 - 2 \left( p_{IZ} + p_{ZI} \right) \right)^{6}
\end{equation}
\begin{equation}
   \approx  \left( 1 - 12 \left(  8\left(\dfrac{ \epsilon}{15}\right) + 4\left(\dfrac{T}{3\tau_{D}}\right)\right) \right).
\end{equation}
This simplifies to,
\begin{equation}
  \approx  1 - \dfrac{96 \epsilon}{15} - \dfrac{16}{3 T\tau_{D}}.
  \label{type_2}
\end{equation}
Therefore, the total type-II error upto first order in $\epsilon$
and $T/\tau_{D}$ using Eqs.~(\ref{type_1}) and (\ref{type_2}) is given by
\begin{equation}
  =  1 - 1032\left(\dfrac{\epsilon}{15}\right) - 160\left(\dfrac{T}{\tau_{D}}\right) - 36(1-\delta).
\end{equation}

\vspace{10pt}

\subsection{Type III: Measurement Errors} 

This accounts for errors during the \(Z\)-measurements on the cluster state qubits. The probability of a \(Z\)-measurement error on cluster state qubits is given by:
\[
p_Z = \frac{2}{3} \epsilon
\]
We categorize the following Pauli errors:

\begin{itemize}
    \item In the case of a \(Z\) error: The qubit remains unaffected, resulting in no error:
    \[
    \alpha |0\rangle - \beta |1\rangle \longrightarrow \text{No error}
    \]
    
    \item In the case of an \(X\) error: The qubit states flip, causing an error:
    \[
    \alpha |1\rangle + \beta |0\rangle \longrightarrow \text{Error}
    \]
    
    \item In the case of a \(Y\) error: The qubit undergoes a phase flip and state flip, leading to an error:
    \[
    \alpha |1\rangle - \beta |0\rangle \longrightarrow \text{Error}
    \]
\end{itemize}
Since there are 6 face qubits to be measured, the total measurement error contribution is:
\[
6 \cdot P_Z = 4 \epsilon \text{ per } \delta q.
\]
The contribution of this error to the expectation value \(\left\langle K_{\partial q}\right\rangle\) is:
\[
(1 - 8 \epsilon).
\]

% This accounts for errors in $Z$ measurements at the cluster state qubits.

% Error in $Z$-measurment at cluster stote qubits $: P_{Z}=\frac{2}{3}\epsilon$.

% We have the follow Pauli errors,

% In case of $Z $ error,
% $$
% \alpha|0\rangle-\beta|1\rangle  \rightarrow \text{No error},
% $$

% In case of X error,
% $$\alpha|1|+\beta|0\rangle \rightarrow \text{error}$$

% In case of Y error: 
% $$\alpha|1\rangle-|0\rangle \rightarrow \text{error}$$

% We have $6$ face qubits to be measured. Thus,
% $\Rightarrow 6 p_{z}=4 \epsilon$ per $\delta q$. The error contribution in  
% $\left\langle K_{\partial q\rangle}\right.$ is $(1-8 \epsilon)$.

\subsection{Threshold Condition}
%  Combining all three categories of errors, we get,

%  \begin{equation}
%      1- \dfrac{1152 \epsilon}{15} - 168\dfrac{T}{\tau_{D}} - 42(1-\delta)
%      \label{thres}
%  \end{equation}
% Using the error threshold criterion for  measurement-based
% quantum computation, Eq.(\ref{thres}) must be greater than $0.7$. This leads to,
% \begin{equation}
%   \epsilon + \frac{35}{16} \frac{T}{\tau_D} + \frac{35}{64} (1 - \sqrt{\beta_{1}}) < 0.0039    
% \end{equation}
By combining all three categories of errors—namely, errors from Bell state creation, teleported CNOT operations, and final qubit measurements—we arrive at the following expression for the total error contribution:
\begin{equation}
1 - \dfrac{1152 \epsilon}{15} - 168 \dfrac{T}{\tau_{D}} - 42(1 - \delta).
\label{thres}
\end{equation}
Here,
\begin{itemize}
    \item \(\epsilon\) represents the gate error.
    \item \(T/\tau_{D}\) is the ratio of the time step duration to the qubit decoherence time, capturing the effects of decoherence on the system.
    \item \(\delta\) is the depolarizing noise parameter.
\end{itemize}

To ensure that the system operates within the fault-tolerant regime, we apply the error threshold criterion for measurement-based quantum computation (MBQC). According to this criterion, the expression in Eq.~(\ref{thres}) must be greater than 0.7, which is the minimum stabilizer expectation value required for fault tolerance. This condition leads to the following inequality:
\begin{equation}
\epsilon + \frac{35}{16} \frac{T}{\tau_D} + \frac{35}{64} (1 - \sqrt{\beta_1}) < 3.9\times 10^{-3}.
\label{ineq}
\end{equation}
Here, $\beta_1$ is a parameter describing depolarizing noise, with \(\sqrt{\beta_1}\) is found to be equal to the depolarizing noise parameter. This inequality shows that, in order to meet the fault-tolerance condition, the combined contributions from gate errors, decoherence effects, and depolarizing noise must remain sufficiently small. Specifically, the sum of the terms in LHS of Eq.(\ref{ineq}) must stay below the threshold value of 0.0039 for the system to operate fault-tolerantly in a measurement-based quantum computation framework.

\nocite{*}

\bibliography{apssamp}% Produces the bibliography via BibTeX.

\end{document}